\documentclass[a4paper,11pt,titlepage,oneside,usenames,dvipsnames,svgnames,x11names,citecolor=NavyBlue,linkcolor=OliveGreen,urlcolor=blue]{article}
\pdfoutput=1
\usepackage{jheppub}
\usepackage{amsfonts, amsmath, amssymb}
\usepackage{latexsym}
\usepackage{breqn}
\usepackage{graphicx}
\usepackage[normalem]{ulem}
\usepackage[usenames,dvipsnames,svgnames,x11names]{xcolor}
\makeatletter
\def\@fpheader{\relax}
\makeatother

\def\al{\alpha}
\def\bt{\beta}
\def\gm{\gamma}

\newcommand{\gp}{\color{black}}
\newcommand{\ab}{\color{black}}

\preprint{ CCTP-2025-15 \\ \hspace*{\fill}ITCP-2025-15}

\title{Energy Reflection and Transmission of Interfaces in $T\bar{T}$-deformed CFT}

\author[1,2]{Avik Banerjee,}
\emailAdd{avik.banerjee@pwr.edu.pl}
\affiliation[1]{Institute of Theoretical Physics, Wroc\l{}aw  University  of  Science  and  Technology,  50-370  Wroc\l{}aw,  Poland}
\affiliation[2]{Crete Center for Theoretical Physics, Institute of Theoretical and Computational Physics,
Department of Physics, University of Crete, 70013 Heraklion, Greece}
\author[3]{and Giuseppe Policastro}
\emailAdd{giuseppe.policastro@phys.ens.fr}
\affiliation[3]{Laboratoire de Physique de l'\'{E}cole Normale Supérieure, ENS, Universit\'{e} PSL, CNRS, Sorbonne Universit\'{e}, Universit\'{e} de Paris, F-75005 Paris, France}

\abstract{ Conformal interfaces gluing a pair of two-dimensional conformal field theories enjoy a large {\gp degree} of universality in terms of the {\ab coefficients} of reflection and transmission of energy, that describe the scattering of conformal matter at the interface. {\ab In this article, we study these coefficients beyond conformality, by gluing a pair of $T\bar T$-deformed 2D CFTs across an interface, which requires the condition  $c_L \mu_L  = c_R \mu_R $ to be obeyed.}  We show that, at least when the interface admits a holographic description, the $T\bar T$ deformation of the CFTs can be extended to the interface. We propose a generalization of the linear matching condition {\ab in the universal sector} of the undeformed ICFT to a non-linear one, which is captured by a universal antisymmetric \emph{transmission function} of the incoming fluxes.  We employ the flow equations of the $T\bar T$-deformed CFTs to compute this function in two special classes of states, namely the non-equilibrium steady state (NESS) and scattering state.  We show that the results can also be reproduced using holographic techniques in the bulk dual of these states. 

}

\begin{document}

\maketitle

\section{Introduction}

The study of Interface Conformal Field Theories (ICFTs) has attracted considerable attention in recent years, with applications ranging from condensed matter physics to string theory. While for certain questions an ICFT can be mapped to a boundary conformal field theory (BCFT) by the folding procedure, there is a growing appreciation for the fact that ICFTs have distinctive features compared to a generic BCFT. These features can be seen, for instance, in the study of the entanglement between two CFTs joined by an interface (characterized by the so-called $g$-factor and the effective central charge) \cite{Karch:2023evr,Karch:2024udk,Afxonidis:2024gne}, and in the transmission of energy between the two CFTs \cite{Quella:2006de,Meineri:2019ycm}. The latter will be the topic of interest for this paper. 

A very remarkable fact about conformal interfaces in 2D CFTs is that the transmission of energy across the interface is \emph{universal} \cite{Meineri:2019ycm}, in the sense that it is independent of the details of the excitations that carry the energy (under some genericity assumptions that we will spell out later). To each interface one can assign a number, the energy transmission coefficient, or  equivalently the reflection coefficient (as there is no absorption at a conformal interface in 2D), which determines the fraction of energy transmitted or reflected at the interface between the two bulk CFTs. In the absence of universality, one would describe the transmission by something like an S-matrix, that would depend on the incoming state, and would contain \emph{a priori} infinitely many parameters. The universality reduces all these parameters to a single one. It also implies that, in the energy-momentum tensor sector, one can consider the conformal interface condition, i.e. the continuity of the energy flux across the interface, as a \emph{linear} condition for the components of the energy-momentum tensor.   

The energy transmission coefficient has been studied for interfaces of free CFTs \cite{Bachas:2001vj}, rational CFTs \cite{Furuta:2025ahl,Makabe:2017ygy}, and holographic CFTs \cite{Bachas:2020yxv, Baig:2022cnb, Bachas:2022etu,Liu:2025khw}. Apart from these few cases, not much is known in general; for instance, we do not know what is the behavior of the transmission coefficient under an RG flow, or under fusion of two interfaces. Part of the difficulty lies in the fact that both of these questions require breaking the conformal symmetry preserved by the interface. We are thus led to consider the following question: how is the energy transmitted across a non-conformal interface? In which way is the universality broken? 

The question is too broad to be answered in full generality. By breaking conformal invariance we lose all the tools that make the study of CFT tractable, and we have to rely on some form of perturbation theory. The only exception, to the best of our knowledge, is the $T\bar T$ deformation \cite{Zamolodchikov:2004ce,Dubovsky:2012wk,Caselle:2013dra,Cavaglia:2016oda,Smirnov:2016lqw}, which allows one to move away from the conformal point and retain the solvability of the undeformed theory. Despite having the standard pathologies of an irrelevant deformation, namely non-locality, lack of UV-completeness, etc., it has gained a significant amount of interest owing to its integrability, factorization property, and holographic control. These remarkable properties have  led to extensive study in the deformed theory in the context of deformed spectrum, S-matrix, entanglement properties, thermal transport properties, etc. (see e.g., the reviews \cite{Jiang:2019epa,He:2025ppz} and references therein). However, the energy transport properties across an interface gluing a pair of deformed theories are yet to be investigated in detail, which is the goal of this paper.\footnote{We should mention that there have been other studies of boundaries and interfaces in $T\bar T$-deformed theory, see \cite{Brizio:2024doe,Wang:2024jem,Dey:2025qms}.}  

To study such energy transport properties, we will primarily make use of the fact that the $T \bar T$ deformation can be described as a dynamical (i.e. state-dependent) coordinate change \cite{Guica:2019nzm}. The effect of the coordinate change can be encoded in a \emph{flow equation} for the expectation value of the energy-momentum tensor, which is all that we need to extract the information of the energy transport. Notice that after we break the conformal symmetry in the bulk CFT, it is not obvious that this determines uniquely the breaking on the interface: one could consider additional deformations localized to the interface; without the constraint of conformal invariance, it is also not clear what conditions should be imposed to characterize the interface. One of the main points of this paper is that the flow equation gives a prescription to  uniquely lift an  interface defined in the undeformed CFT to the deformed theory. It is conceivable that the deformed theory might admit also other interfaces not determined by this procedure; we will not investigate this question. 

Since the flow equation is state-dependent, as mentioned above, in order to carry out the procedure explicitly we have to choose a particular state. We found that there are two cases that we can analyze explicitly. One is the NESS (non-equilibrium steady-state) created by taking two BCFTs at different temperatures, joining them at some initial time, and looking at the state created at late times (in the absence of the interface, this would be simply a boosted thermal state, and its $T \bar T$-deformation was studied in \cite{Medenjak:2020bpe,Medenjak:2020ppv}).
The second is a ``scattering" state made of small plane-wave excitations around the vacuum. This case can be analyzed in a perturbative expansion in the amplitude of the fluctuations. 

After computing the energy transmission for these two states using the flow method, as a consistency check of our results, we shall reproduce them from the holographic computations in the bulk dual of these states. The $T \bar T$ deformation is  known to admit a relatively simple holographic dual description, in terms of a finite cutoff \cite{McGough:2016lol} or modified boundary conditions \cite{Guica:2019nzm}. In the case of pure 3D gravity the two  prescriptions coincide, and it can be shown that they directly yield the flow equations \cite{Guica:2019nzm}.   By analogy with the calculations of energy transmission in holographic CFT models, where the interface is described by a brane in the bulk AdS space \cite{Bachas:2020yxv,Bachas:2021tnp}, we shall also consider gluing a pair of deformed geometries across a thin brane, within the modified boundary condition picture. We shall show that the holographic results match precisely with the results obtained with the flow method. Notice that, this is a non-trivial check of the prescription, since the presence of a brane introduces a source of matter in the bulk.

{\gp The main result of this paper is the precise realization of the modified energy transport properties induced by the deformation. We have to distinguish the two cases. In the case of the NESS, we found that the linear matching conditions of the stress-tensor vevs \cite{Meineri:2019ycm} across the conformal interface are generalized to non-linear conditions, wherein the single transmission coefficient becomes a \emph{transmission function}, which is an antisymmetric function of the incoming energy fluxes on the interface. This function contains an infinite number of non-linear transmission coefficients, which can be considered universal in the same sense as the CFT coefficient: the amount of reflected and transmitted energy is only a function of the incoming fluxes and not of the way the energy is distributed in different modes of the theory.  However, within our techniques, we cannot show that the universality holds in the sense of the independence on the operator used to create the excitation.
We cannot find a closed form for the transmission function, but we can evaluate it to arbitrarily high order in the incoming fluxes. In the case of the scattering state, the time dependence introduces additional complications that make us restrict our analysis to linear order. For the scattering of single frequency states, we can define a frequency-dependent transmission coefficient, which however takes physical values only in the regime $\omega^2 {\ab \leq} \frac 6 {c\mid\mu\mid}$. For states that contain waves of different frequencies, the matching can be expressed as a function of the time derivatives of the incoming fluxes. 
Finally, within both the methods, we observe that the gluing of two deformed CFTs with central charges $c_{L,R}$ and deformation parameters $\mu_{L,R}$ must obey the condition $c_L \mu_L =c_R \mu_R $, which we conjecture to remain true in general. This condition has a very natural interpretation in the finite cutoff holographic prescription: it simply says that the radial cutoff must match on the two sides.}

The paper is organized as follows. In section \ref{sec:CFT-universality}, we provide the necessary review of the universality of the reflection and transmission coefficients in two-dimensional interface CFTs. In section \ref{sec:floweq}, we review the flow equations of $T \bar T$ deformed  CFTs and their holographic interpretation as mixed boundary conditions. In section \ref{sec:coefficientsfloweq}, we discuss the non-linear matching conditions for the  deformed ICFT and demonstrate our proposal of computing the transmission function  in the context of NESS and scattering state, using the flow equations of the deformed theory. In sections \ref{sec:holoNESS} and \ref{sec:Gravscatt}, we reproduce the field-theoretic results for our new measure of the energy transmission using holographic techniques in the bulk dual of these states. Finally, in section \ref{sec:conclusion}, we summarize our results and conclude with possible future directions of our work.

\section{The universality of energy transmission in 1+1 dimensions}\label{sec:CFT-universality}

\begin{figure}[h!]
    \centering
    \includegraphics[scale=0.4]{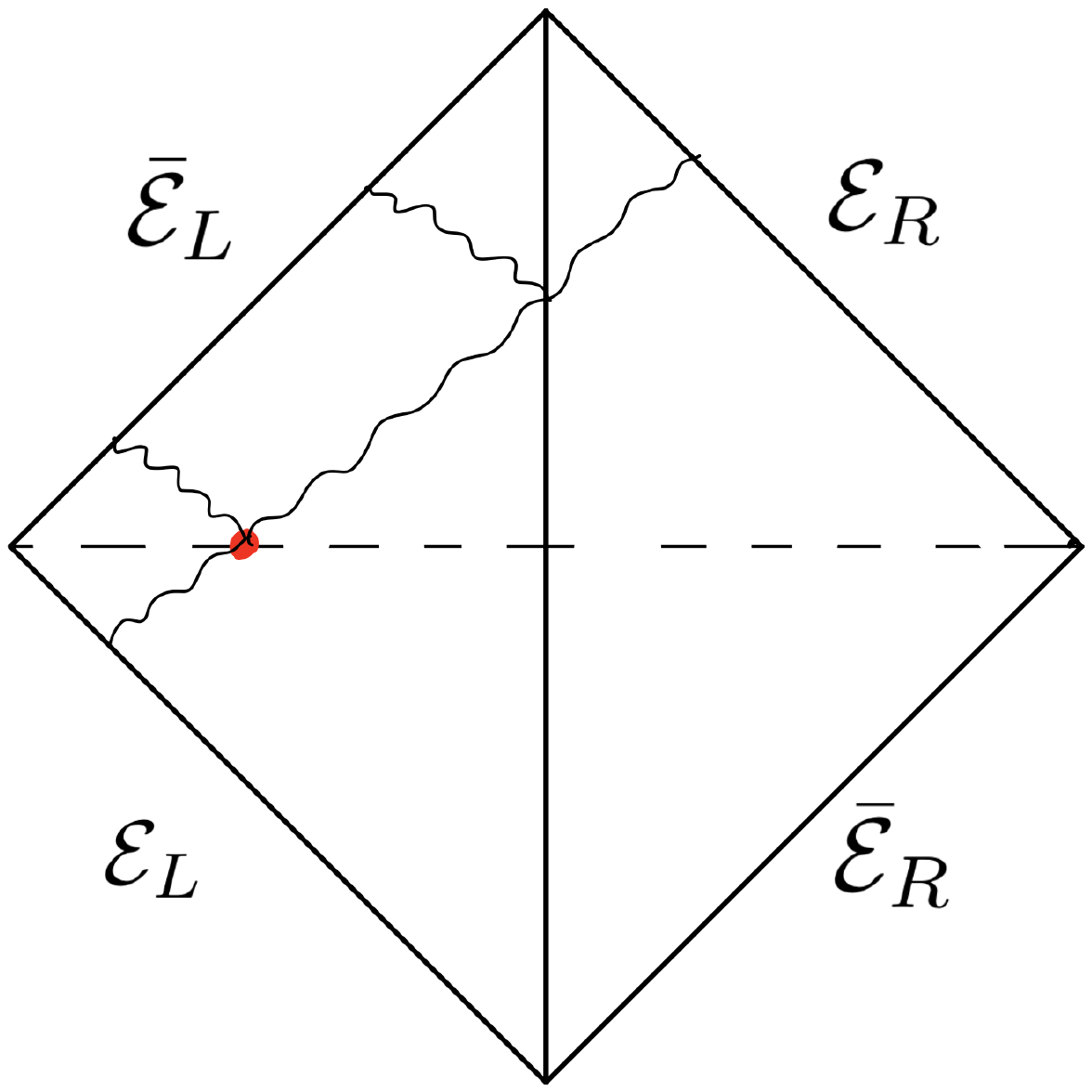}
    \caption{Conformal diagram of flat space with a timelike defect separating it into two halves. The figure demonstrates the  incident, reflected and transmitted excitation due to the action of a local operator in CFT$_L$. The fluxes are collected at various null infinities. }
    \label{Figure2}
\end{figure}

In this section, we shall briefly summarize the universality of the reflection and transmission coefficients of energy across a conformal interface gluing a pair of two-dimensional CFTs, following \cite{Meineri:2019ycm}. 
Consider an interface gluing two 1+1 dimensional CFTs with central charges $c_L$ and $c_R$ respectively. Using translational invariance we shall fix the interface at $x=0$.  The interface can be thought of as the worldline of the junction of two semi-infinite quantum wires glued at $x=0$. The residual symmetries are then the ones that leave the interface invariant, namely, translation and special conformal transformation along time, and dilatation. In this setup, we want to measure the coefficients of reflection and transmission of energy associated with the interface, which are given by  
\begin{align}
    \mathcal R=\frac{\text{Reflected energy}}{\text{Incident energy}},~~~~~ \mathcal T=\frac{\text{Transmitted energy}}{\text{Incident energy}},~~~~~  \mathcal R +\mathcal T =1\, ,
\end{align}
where the last equation is simply restating the fact that a conformal interface does not absorb energy and hence, any incident excitation only gets reflected and transmitted with the coefficients adding up to unity due to energy conservation. These equations hold for either side of the interface and the energies will be measured at various null infinities as illustrated in   Figure \ref{Figure2}. 
Once the ideas are defined, we need to quantify the measures of the coefficients by suitably choosing the operators and states.  In two-dimensional CFTs, the stress tensor enjoys holomorphic splitting and the energy density at any point decomposes into a right-moving and a left-moving component
\begin{equation}
    \left\langle T^{00}(x,t)\right\rangle=\left(\left\langle T(u)\right\rangle +\left\langle \bar T(v)\right\rangle\right),
\end{equation}
with $u=x-t$ and $v= x+t$. This implies that the left and the right movers blindly cross each other without any exchange of energy and this helps to a great extent to accurately measure flux along the two null directions at the infinities. The two operators relevant for the measurement of these fluxes are the ANEC operators given by:
\begin{align}
    {\cal{E}} = \int_{-\infty}^{+\infty} T(u) du~,~~~\bar {\cal{E}} = \int_{-\infty}^{+\infty} \bar T(v) dv.
\end{align}
These are special kinds of the light-ray operators and in this context, they simply coincide with the null components of the momentum operators, $ {\cal{E}} =P_u$ and $\bar  {\cal{E}} = P_{v} $. Hence in any state, they will have non-negative eigenvalues and are therefore suitable for the measurement of energy fluxes.

Now since the holomorphic splitting is a local property, this holds even in the presence of an interface. However, the stress tensors of the two CFTs must satisfy the gluing condition at the interface

\begin{equation}
    T_L(u=-t)-\bar T_L(v=t)= T_R(u=-t)-\bar T_R(v=t),
\end{equation}  
which simply follows from the invariance under time translation and dilatation,  implying the continuity of the energy flux across the  interface in the absence of any absorption. Two trivial solutions of this gluing equation are:
\begin{align}
  \nonumber  T_L(u=-t) =&~ T_R(u=-t) ~,~~ \bar T_L(v=t) = \bar T_R(v=t) ,~~~~~~~~~~~\text{Topological interface,}\\
\nonumber    T_L(u=-t) =&~ \bar T_L(v=t) ~,~~ T_R(u=-t) = \bar T_R(v=t) ,~~~~~~~~~~~\text{Factorizing interface}.
\end{align}
The first case corresponds to $\mathcal T_{L,R}=1$, $\mathcal R_{L,R}=0$, while the latter corresponds to $\mathcal T_{L,R}=0$, $\mathcal R_{L,R}=1$. However, in the following we shall choose a generic interface with $0< \mathcal T_{L,R}<1$, $0< \mathcal{R}_{L,R}<1$ with $\mathcal{T}_{L(R)}+\mathcal{R}_{L(R)}=1$.

Having identified the suitable operators, next we look for asymptotic states. For precise measurement of these coefficients, we need to decouple various fluxes from each other, which is however difficult in CFTs due to the power-law tails. To circumvent this, the smearing function used to delocalize the excitations created by local operators are chosen to have a finite support. Also, the excitations are typically prepared and measured far away from interface. In what follows, we shall be interested in a situation where the excitation scattered against the interface will be prepared in CFT$_L$. In that case, the asymptotic states are defined as 
\begin{equation}
    \mid O_{L},D \rangle_{I}= \int du dv ~ f(u)f(v +D) O_{L}(z,\bar z) \mid 0\rangle_{I}.
\end{equation}
Here $O_L$ is a generic local operator belonging to CFT$_L$  and the subscript $I$ on a state denotes that it belongs to the Hilbert space of defect CFT containing an impurity at $x=0$. The wavepacket is chosen to have finite support, i.e.
$$\int_{-\infty}^{+\infty} f(x)dx=1~,~~~~ f(x)=0 ~ \forall \mid x\mid >l~,$$
and $D$ denotes the distance from the interface along the  right-moving null direction. {\gp We will be interested in the limit  $D \rightarrow \infty$; by taking also $l \rightarrow \infty$, while keeping $l << D$, we can consider plane-wave excitations that we will use later}.

Having defined the states and the observables, we now finally define the various coefficients as follows

\begin{align}
    \mathcal T_L =& \lim_{D \rightarrow \infty} \frac{\left\langle O_L,D \mid  {\cal{E}}_R \mid O_L,D \right\rangle_I}{\left\langle O_L,D \mid  {\cal{E}}_L \mid O_L,D \right\rangle}, \label{Tl}\\
     \mathcal R_L =& \lim_{D \rightarrow \infty} \frac{\left\langle O_L,D \mid \bar  {\cal{E}}_L \mid O_L,D \right\rangle_I- \left\langle O_L,D \mid \bar  {\cal{E}}_L \mid O_L,D \right\rangle}{\left\langle O_L,D \mid  {\cal{E}}_L \mid O_L,D \right\rangle}, \label{Rl}\\
      \mathcal T_R =& \lim_{D \rightarrow \infty} \frac{\left\langle O_R,D \mid \bar  {\cal{E}}_L \mid O_R,D \right\rangle_I}{\left\langle O_R,D \mid \bar  {\cal{E}}_R \mid O_R,D \right\rangle},\label{TR}\\
      \mathcal R_R =& \lim_{D \rightarrow \infty} \frac{\left\langle O_R,D \mid   {\cal{E}}_R \mid O_R,D \right\rangle_I- \left\langle O_R,D \mid   {\cal{E}}_R \mid O_R,D \right\rangle}{\left\langle O_R,D \mid \bar  {\cal{E}} _R \mid O_R,D \right\rangle}. \label{RR}
\end{align}
Here $\mathcal T_{L(R)} , \mathcal{R}_{L(R)}$ measures transmission and reflection coefficients when the excitation is incident on the interface from left (right). Similarly, the subscripts on the various operators indicate whether they belong to CFT$_L$ or  CFT$_R$. The states without the subscript $I$ are analogous to the ones with the subscript, except that they are created on top of the translational invariant vacuum. In computing the reflection coefficients we have also subtracted the contribution that reaches infinity directly without being reflected from the interface.

Now the conservation law $\mathcal T_{L(R)}+\mathcal R_{L(R)}=1$ (see\cite{Meineri:2019ycm} for a detailed proof of this relation starting from the definitions \eqref{Tl}-\eqref{RR}) leaves only two of the coefficients independent, which we choose to be $\mathcal{T}_{L,R}$. In what follows, we provide a  derivation of  the transmission coefficients under the simplifying assumption that the operator creating the excitation is holomorphic.
For the proof with a generic operator, we refer to \cite{Meineri:2019ycm}. 

To evaluate $\mathcal T_L$, we need to compute the three-point function\footnote{We switch to euclidean signature for the derivation.} $\left\langle O_L^1(z_1)  T_R(z) O_L^2(z_2)\right\rangle_I $. Here $O_L^1, O_L^2$ are local operators in CFT$_L$ which we shall initially choose to be linear combinations of quasi-primaries.  Note that for states created using holomorphic operators in CFT$_L$, there is no dependence on the position of the wavepacket along $\bar z$ and the $D \rightarrow \infty$ limit can be dropped. Now, we shall first consider the OPE expansion of the two $O_L$s. Then the three-point function is essentially given by the sum of two-point functions of holomorphic operators of CFT$_L$ with $T_R$. Now the two-point function of a holomorphic operator with a generic operator in the presence of the interface is given by \cite{Meineri:2019ycm}
\begin{align}
    \left\langle O_1 (z_1,\bar z_1)  O_2(z_2)\right \rangle = \frac{c_{12}}{(z_1-z_2)^{h_1+h_2-\bar h_1}(z_1+\bar z_1)^{h_1+\bar h_1- h_2}(z_2+\bar z_1)^{h_2+\bar h_1- h_1}}.
\end{align}
If the operators are on the opposite sides of the interface, then $c_{12} \neq 0$ if and only if $h_1 -\bar h_1 -h_2 =0$. Now if $O_1$ is purely antiholomorphic, then non-vanishing of the two point function will require $\bar h_1 +h_2=0 $, which is not possible in unitary CFTs. On the other hand, if $O_1$ is purely holomorphic, we must have 
\begin{align}
    \left\langle O_1 (z_1,)  O_2(z_2)\right \rangle = \frac{c_{12}~ \delta_{h_1,h_2}}{(z_1-z_2)^{h_1+h_2}}
\end{align}
So holomorphic operators only with equal weight correlate across the interface. Now under the assumption that  the  CFT$_L$ has only Virasoro symmetry and not a larger one, there is a unique spin 2 holomorphic operator in CFT$_L$, which is $T_L$. Then from \eqref{Tl}, we have 
\begin{align}
    \mathcal T_L= \frac{c_{LR}}{c_L},
\end{align}
where $c_{LR}$ is defined by 
\begin{align}
\nonumber \left\langle T_L(z_1)T_R(z_2)\right\rangle=\frac{c_{LR}/2}{(z_1-z_2)^4} .
\end{align}
A similar computation for $\mathcal T_R$ would have led to $\mathcal T_R= c_{LR}/{c_R}$.  Thus, the coefficients are completely fixed by the central charges of the two theories and by $c_{LR}$ \cite{Quella:2006de} and are completely insensitive to the local operators used to create the states as well as to the choice of the wave packets. Note that, in addition to the conservation law, the transmission coefficients additionally obey $c_L \mathcal T _L= c_R \mathcal T_R$, which is the condition of detailed balance. This leaves only one out the four coefficients \eqref{Tl}-\eqref{RR} independent. Finally note that, since the kinematic factors under the integral drop out between the numerator and the denominator, the above analysis will go through even if we take derivatives of the three-point function w.r.t $z_1, z_2$ or multiply with $z,z^2$. This tells us that the above results are true for generic local operators in CFT$_L$ and also for the measurement of charges associated with dilatation or special conformal transformation. 


\section{ $T\bar{T}$ deformation: flow equations and mixed boundary condition}\label{sec:floweq}
The $T\bar{T}$ deformation is a universal irrelevant deformation of any local two-dimensional quantum (conformal) field theory, defined by the differential relation
\begin{equation}
    \partial_{\mu}S^{[\mu]}= -\frac{1}{2} \int~ d^2x \sqrt{\gamma} \left(T^{\alpha\beta}T_{\alpha\beta}-\Theta^2\right), \label{defttbar}
\end{equation}
where $\Theta = T_\alpha^\alpha$, and the quantities appearing on the right-hand side are evaluated in the deformed theory.  The deformation is parametrized by $\mu$ which has dimensions length$^{2}$, rendering it irrelevant and the  deformed theory is known to be non-local. Since our discussions will primarily revolve around the vev of the stress-tensor, we shall restrict ourselves to scales much larger than the scale of non-locality  $\sqrt{\mu}$, so that the deformed theory is quasi-local and the stress tensor is unambiguously defined in the usual sense.

Given the fact that $T\bar{T}$  is a double-trace deformation where the deforming operator also depends on the source, it is natural to expect the holographic dictionary for the deformed theory to be modified. In fact, such deformations are known to lead to mixed boundary conditions. In the context of $T\bar{T}$ deformation, the deformed dictionary can be most efficiently worked out using the variational principle \cite{Guica:2019nzm}, which we briefly outline here. Taking a variation of the defining relation \eqref{defttbar} with respect to the metric and after some simple algebra, one arrives at,
\begin{align} \label{fe1}
\nonumber \partial_{\mu}\left(\sqrt{\gamma}  T_{\al\bt}\right) \delta \gamma^{\al\bt} +&\, \sqrt{\gamma} T_{\al\bt} \partial_{\mu} \left(\delta \gamma^{\al\bt}\right) \\
	 =&\ \sqrt{\gamma}\left[\left(-\frac{1}{2} \gamma_{\al\bt} \mathcal{O}_{T\bar{T}}-2 T_{\al\gamma} T_{\bt}^{\gm} +2 \Theta T_{\al\bt}\right) \delta \gamma^{\alpha \beta}+ 2 T_{\al\bt} \delta \left(T^{\al\bt} - \gamma^{\al\bt} \Theta \right)\right].  \nonumber
\end{align}
where we have used $$\delta S^{[\mu]}=
\left(\frac{1}{2} \int d^2 x~ \sqrt{\gamma}~T_{\alpha\beta}\delta\gamma^{\alpha\beta}\right)^{[\mu]},$$ and $T_{\al\bt}$ stands for its expectation value for the rest of the section.
Comparing the quantity being varied and the coefficient of the variation, one arrives at the flow equations
\begin{equation} \label{fe2}
\partial_{\mu} \gm^{\al\bt} = 2 \left(T^{\al\bt} -\gm^{\al\bt} T\right), \qquad \partial_{\mu}\left(\sqrt{\gamma} T_{\al\bt}\right) =\sqrt{\gm}( 2 T T_{\al\bt} - 2 T _{\al\gm}T_{\bt}^{\gm} -\frac{1}{2}\gamma_{\al\bt}\mathcal{O}_{T\bar{T}}),
\end{equation}
where $\mathcal{O}_{T\bar{T}}= T^{\alpha\beta}T_{\alpha\beta}-\Theta^2$. Upon introducing $\hat{T}_{\al\bt}= T_{\al\bt}- \gm_{\al\bt} \Theta$, these equations can be written in a more compact form as
\begin{align}
\partial_{\mu} \gm_{\al\bt} =&\ -  2 \hat{T}_{\al\bt},  \label{fe21} \\
\partial_{\mu} \hat{T}_{\al\bt} =&\  -\hat{T}_{\al\gm} \hat{T}_{\bt}^{\gm}. \label{fe22}
\end{align}
where in arriving at (\ref{fe22}) we have used the relations $\partial_{\mu}\sqrt{\gm} = \sqrt{\gm} \Theta$, $\Theta T_{\al\bt} -T_{\al\gm}T_{\bt}^{\gm}= -\frac{1}{2} \gm_{\al\bt} \mathcal{O}_{T\bar{T}}$, and $\partial_{\mu} \Theta = -T^{\al\bt} T_{\al\bt}$ (see Appendix A of \cite{Guica:2019nzm, Banerjee:2024wtl}). By differentiating (\ref{fe21}) twice and using (\ref{fe22}) we arrive at the final form of the flow equations
\begin{align} 
\partial_{\mu}^3 \gm_{\al\bt} =&\  0,  \label{fe31} \\
\partial_{\mu} \hat{T}_{\al\bt} =&\  -\hat{T}_{\al\gm} \hat{T}_{\bt}^{\gm}. \label{fe32}
\end{align}
These equations can now  be easily solved to get the deformed metric and stress tensor expectation value as
\begin{align} \gm^{[\mu]}_{\al\bt} =&\  \gm^{[0]}_{\al\bt} - 2 \mu  \hat{T}^{[0]}_{\al\bt} +\mu^2  \hat{T}^{[0]}_{\al\rho} \gamma^{[0]\rho\sigma}\hat{T}_{\sigma \bt}^{[0]} , \label{solfe1}\\
 \hat{T}^{[\mu]}_{\al\bt} =&\  \hat{T}^{[0]}_{\al\bt} -\mu \hat{T}^{[0]}_{\al\rho} \gamma^{[0]\rho\sigma}\hat{T}_{\sigma \bt}^{[0]}.  \label{solfe2}
\end{align}
Note that these solutions are non-perturbative in the deformation parameter. {\gp It is also useful to know that as a consequence of the flow equations, the following trace relation holds:\footnote{In \cite{Guica:2022gts} it is argued that this relation should be valid also at the quantum level up to total derivative terms, which were found to vanish at linear order in the perturbation.} 
\begin{equation}
    \Theta = - \mu \, \mathcal{O}_{T\bar{T}} \,.
\end{equation}
}
Let us now plug back the undeformed dictionary in \eqref{solfe1}. In the case of pure gravity, identifying
\begin{equation}
    \gamma^{[0]}_{\alpha\beta} \equiv g^{(0)}_{\alpha \beta}~,~~ \hat{T}^{[0]}_{\alpha \beta}\equiv  \frac{1}{8\pi G l}  g^{(2)}_{\alpha \beta}~, ~~\hat{T}^{[0]}_{\al\rho} \gamma^{[0]\rho\sigma}\hat{T}_{\sigma \bt}^{[0]} \equiv \frac{1}{(4\pi G l)^2} g^{(4)}_{\alpha \beta} ,  \label{undefmap}
\end{equation}
with the coefficients of the Fefferman-graham expansion for the solution of Einstein's gravity with $\Lambda = -2/l^2$,
$$ds^2 = l^2 \frac{d\rho^2}{\rho^2}+ \left(\frac{g^{(0)}_{\alpha \beta}}{\rho}+g^{(0)}_{\alpha \beta}+ \rho g^{(4)}_{\alpha \beta}\right) dx^{\alpha} dx^{\beta},$$ 
we readily note that fixing the deformed metric in \eqref{solfe1} now amounts to imposing Dirichlet boundary condition at a finite radial slice $\rho_c= -\frac{\mu}{4\pi G l}$. In this case, one can further show that the deformed stress tensor now has the identification 
\begin{equation*}
    {T}^{[\mu]}_{\al\bt}= T^{BY}_{\al\bt}(\rho_c)-\frac{g_{\al\bt}(\rho_c)}{8\pi Gl}.
\end{equation*}
However, both of these identifications are true only for the particular sign of the deformation parameter and in the absence of any matter. In the computations to follow, we shall be interested in both signs of the deformation and also consider matter minimally interacting with gravity. In this case, the last of the identifications made in \eqref{undefmap} does not work rendering the finite cut-off prescription invalid. However, just from the first two identifications in \eqref{undefmap}, we can infer that fixing $\gm^{[\mu]}$ can be thought of as a mixed boundary condition at the original boundary instead of a Dirichlet boundary condition on $g^{(0)}$. For the rest of the discussions, we shall stick to this interpretation of the deformation. 

\subsection{ Deformed Ba\~nados spacetime}
In this section we shall provide the details of the deformed spacetimes which will essentially set the stage for the holographic computations of the coefficients.  Towards that, we begin with seeking the most general bulk spacetime that gives rise to some fixed deformed boundary metric, say $\gm^{[\mu]}_{\al\bt}=\eta_{\al\bt}$, in some coordinates $(U,V)$. Now, since $\sqrt{\gm}R$ is invariant along the flow, for a Ricci flat deformed metric, the undeformed metric must also be Ricci flat, which we can again choose to be $\eta_{\al\bt}$ in some different coordinates, say $(u,v)$. In the undeformed case, for a flat boundary metric $g^{(0)}_{\al\bt}=\eta_{\al\bt}$, the most general solution is the Ba\~nados spacetime characterized by two chiral functions $g^{(2)}_{uu}\equiv \mathcal L(u)~,g^{(2)}_{vv}\equiv \bar{\mathcal L}(v)$. Putting all these back in \eqref{undefmap}, from \eqref{solfe1} we have
 \begin{equation}
     ds^2_{[\mu]}= \left(du - \frac{\mu}{4\pi G l}\bar{\mathcal L}(v) dv\right)\left(dv - \frac{\mu}{4\pi G l}{\mathcal L}(u) du\right). \nonumber
 \end{equation}
 Note that this deformed metric can be cast into a flat metric $ds^2_{[\mu]}=dU dV$ with the help of the pseudoconformal map
 \begin{equation} \label{psc1}
     U= u - \frac{\mu}{4\pi G l}\int^v\bar{\mathcal L}(v') dv'~~,~~~ V= v - \frac{\mu}{4\pi G l}\int^u {\mathcal L}(u') du'.
 \end{equation}
 Note that, this map is invertible exactly when the vevs are constants, or perturbatively, when the vevs are small. For the constant vevs, for example,  the map can be inverted to give
 \begin{equation}
     u= \frac{U+ \frac{\mu}{ 4 \pi G l}\bar{\mathcal L} ~V}{1-\frac{\mu^2}{16 \pi^2 G^2 l^2}{\mathcal L} \bar{\mathcal L}}~~,~~~~ v= \frac{V+ \frac{ \mu}{ 4 \pi G l}{\mathcal L} ~U}{1-\frac{\mu^2}{16 \pi^2 G^2 l^2}{\mathcal L} \bar{\mathcal L}}~. ~~ \label{pcfmap}
 \end{equation}
Finally, the spacetime asymptoting to the flat deformed metric at the boundary can now be easily found in the Fefferman-Graham gauge by using the map \eqref{pcfmap} in the FG expansion of the undeformed spacetime, which gives
\begin{equation}		\label{defbanados2}
ds^2_{[\mu]} = l^2\frac{d\rho^2}{4\rho^2}+\frac{ds^{(0)2}_{[\mu]}}{\rho} + ds^{(2)2}_{[\mu]}+ \rho ~ds^{(4)2}_{[\mu]}~, \nonumber
\end{equation}
where,
\begin{align}
ds^{(0)2}_{[\mu]}=&\  \frac{\left(dU + \frac{\mu}{ 4 \pi G l}\bar{\mathcal L} dV\right)\left(dV + \frac{\mu}{4 \pi G l}{\mathcal L} dU\right)}{\left(1-\frac{\mu^2}{16 
\pi^2 G^2l^2}{\mathcal L} \bar{\mathcal L}\right)^2} , \nonumber  \\
ds^{(2)2}_{[\mu]}= &\ \frac{\left(1+ \frac{\mu^2}{16 
\pi^2 G^2l^2} {\mathcal{L}} \bar{\mathcal L}\right)\left({\mathcal L} dU^2+ \bar{\mathcal L}dV^2\right)+\frac{\mu}{4 \pi G l}{\mathcal L}\bar{\mathcal L}~dUdV}{\left(1-\frac{\mu^2}{16 
\pi^2 G^2l^2}{\mathcal L} \bar{\mathcal L}\right)^2}, \nonumber \\
ds^{(4)2}_{[\mu]}=&\  {\mathcal L} \bar{\mathcal L} ds^{(0)2}_{[\mu]}. \nonumber 	
\end{align}
One can now easily check that
$$ds^{(0)2}_{[\mu]}-\frac{2\mu}{l}ds^{(2)2}_{[\mu]}+ \frac{4\mu^2}{l^2}ds^{(4)2}_{[\mu]}=dUdV.$$

We conclude the section by reporting the expectation value of the  deformed stress tensor in the $(U,V)$ coordinates  
\begin{align} \label{defT}
    T^{[\mu]}_{\al\bt} =\frac{1}{8\pi G l\left(1-\frac{\mu^2}{16 \pi^2 G^2 l^2}{\mathcal L} \bar{\mathcal L}\right)}\begin{pmatrix}
{\mathcal L} & -\frac{\mu}{4 \pi G l} {\mathcal L} \bar{\mathcal L}\\
-\frac{\mu}{4\pi G l} {\mathcal L} \bar{\mathcal L} & \bar{\mathcal L}\\
    \end{pmatrix}.
\end{align}
which can be obtained by starting from the undeformed expectation values $\hat{T}^{[0]}_{uu}= \frac{{\mathcal L}}{8\pi G l}$, $\hat{ T}^{[0]}_{vv}= \frac{\bar {\mathcal L}}{8 \pi G l}$ and using the pseudoconformal map \eqref{pcfmap}.

\section{Energy transmission from the flow equation} \label{sec:coefficientsfloweq}

In this section, we will show that we can deduce unambiguously the modification of the energy transport properties of a conformal interface under a $T \bar T$ deformation, using the description of the deformation as a dynamical change of coordinates, encoded in the flow equations.  

Let us recall for convenience the solution of the flow equations \eqref{solfe1}-\eqref{solfe2} relating the deformed metric and stress-energy tensor  to the undeformed ones :
\begin{align} \gm^{[\mu]}_{\al\bt} =&\  \gm^{[0]}_{\al\bt} - 2 \mu  \hat{T}^{[0]}_{\al\bt} +\mu^2  \hat{T}^{[0]}_{\al\rho} \gamma^{[0]\rho\sigma}\hat{T}_{\sigma \bt}^{[0]} , \\
 \hat{T}^{[\mu]}_{\al\bt} =&\  \hat{T}^{[0]}_{\al\bt} -\mu \hat{T}^{[0]}_{\al\rho} \gamma^{[0]\rho\sigma}\hat{T}_{\sigma \bt}^{[0]}.  
\end{align}
Starting from a flat undeformed metric $ds^2 =  du dv$, and a stress-energy tensor $\hat T^{[0]} = {\mathcal L}(u)du^2+\bar {\mathcal L}(v) dv^2$, we find that the deformed metric is flat in new coordinates $(U,V)$ provided 
\begin{equation}\label{pcfloweq}
    U_{i} = u_{i} - 2  \mu_{i} \,  \bar h_{i}(v_{i}) \,, \quad  V_{i} = v_{i} - 2  \mu_{i} \, h_{i}(u_{i}) \,.
\end{equation}
where $i=L,R$ and we have chosen ${\mathcal L_{i}}(u_{i}) = h'_{i}(u_{i}), \bar {\mathcal L}_{i}(v_{i}) = \bar h'_{i}(v_{i})$ for convenience. Now, for the defect to remain vertical in the $(U_i,V_i)$ coordinates, it cannot remain so in the $(u_i,v_i)$ coordinates of the undeformed ICFT for non-zero vevs of the stress-tensor. To relate  the results to undeformed ICFT  with a vertical interface, we supplement \eqref{pcfloweq} further with a conformal transformation 
\begin{equation}\label{tildemap}
  \tilde u_i = u_i + 2  \mu_i \, \alpha_i(u_i) \,, \quad 
  \tilde v_i  = v_i + 2  \mu_i \, \bar \alpha_i(v_i) \,, 
\end{equation}
claiming that in $(\tilde u_i, \tilde v_i)$ coordinates, the interface is vertical at $\tilde x_i= (\tilde u_i +\tilde v_i)/2=0$. Now, the requirement that the interface is vertical in  the $(U_i,V_i)$ coordinates leads to the constraint
\begin{equation}\label{bc1}
    (h_i(u_i)+ \bar h_i(v_i)+\alpha_i(u_i)+ \bar \alpha_i(v_i))|_{bdy} =0 \,.
\end{equation}
 Note that this condition should hold separately for CFT$_L$ and CFT$_R$. Additionally, we require the time to remain continuous across the interface, that is, 
$V_L-U_L=V_R-U_R$. Now since $U_i=-V_i$ at the interface, the continuity of time across the interface leads to{\footnote{Or equivalently, $\mu_L\left(\bar h_L(v)+  \alpha_L(u)\right)- \mu_R\left(\bar h_R(v)+  \alpha_R(u)\right)\mid_{bdy}=0$.}}
\begin{equation}\label{bc2}
    \mu_L\left(h_L(u)+ \bar \alpha_L(v)\right)- \mu_R\left(h_R(u)+ \bar \alpha_R(v)\right)\mid_{bdy}=0.
\end{equation}
To solve \eqref{bc1}-\eqref{bc2}, we shall first use the conformal map \eqref{tildemap} in the arguments of the functions and then set $\tilde v_i= -\tilde u_i$, with $\tilde u_L=\tilde u_R$ . Thus, these constraints fix only three out of the four functions $\alpha_{L,R}, \bar \alpha_{L,R}$, leaving one unspecified. However, it turns out that the result does not depend on it (in fact this corresponds to a freedom of global time redefinition in the CFT). Note that with this boundary conditions imposed, the corrdinates $(U,V)$ and $(\tilde u, \tilde v)$ are continuous across the (vertical) interface, while $(u,v)$ are discontinuous across it.

Let us take the stress-energy tensor in the $(\tilde u, \tilde v)$ coordinates with the form\footnote{Note that in the undeformed CFT, $\hat T$ and $T$ are  {\gp the same}.} $T_i^{[0]} = f'_i(\tilde u) d \tilde u^2+ \bar f '_i(\tilde v) d \tilde v^2$. Now under a conformal transformation, the stress-energy tensor transforms as 
\begin{equation}\label{EMtransf}
 \left(\frac{d\tilde u}{du_i}\right)^2 T_{\tilde u \tilde u(i)}(\tilde u) =T_{uu(i)}(u_i) + \frac{c_i}{12} \{\tilde u,u_i\} \,.
\end{equation}
Then in the $(u_i,v_i)$ coordinates, the vevs are given by
\begin{equation}\label{eq:f}
   h'_i(u_i)= f'_i(\tilde u) (1+2 \mu_i \alpha'_i(u_i))^2 - \frac {c_i}{12} \{\tilde u, u_i\} \,, \quad \bar h'_i(v_i)= \bar f_i'(\tilde v)(1+2 \mu_i \bar \alpha'_i(v_i))^2 - \frac{c_i}{12}\{\tilde v, v_i\} \,.
\end{equation}
Then using the flow equations, we find the deformed stress-tensor components as
\begin{equation}
    \begin{split}\label{defvev}
        T_{UU(i)} & = \frac{h'_i(u_i)}{1-4 \mu_i^2 h'_i(u_i) \bar h'_i(v_i)} \,, \\
        T_{VV(i)} & = \frac{\bar h'_i(v_i)}{1-4 \mu_i^2 h'_i(u_i) \bar h'_i(v_i)} \,, \\
        T_{UV(i)} & = \frac{-2 \mu_i ~h_i'(u_i) \bar h'_i(v_i)}{1-4 \mu_i^2 h'_i(u_i) \bar h'_i(v_i)} \,.
    \end{split}
\end{equation}
One can check that the trace relation is satisfied:
\begin{equation}\label{trace-eq}
      T_{UV(i)} + 2 \mu_i T_{UU(i)} T_{VV(i)}   - 2 \mu_i T_{UV(i)}^2 = 0 \,.
\end{equation}
Now that we have the stress-tensor vevs in the deformed theory in flat space with a straight interface, we are now ready to discuss the transmission of energy. However, our discussion so far has been rather abstract. In what follows, we shall consider two specific classes of states to demonstrate the method.  

\subsection{Non-equilibrium Steady State (NESS)}
\begin{figure}[tbp]
    \centering
    \begin{minipage}{.48\textwidth}
        \centering
        \includegraphics[scale=0.3]{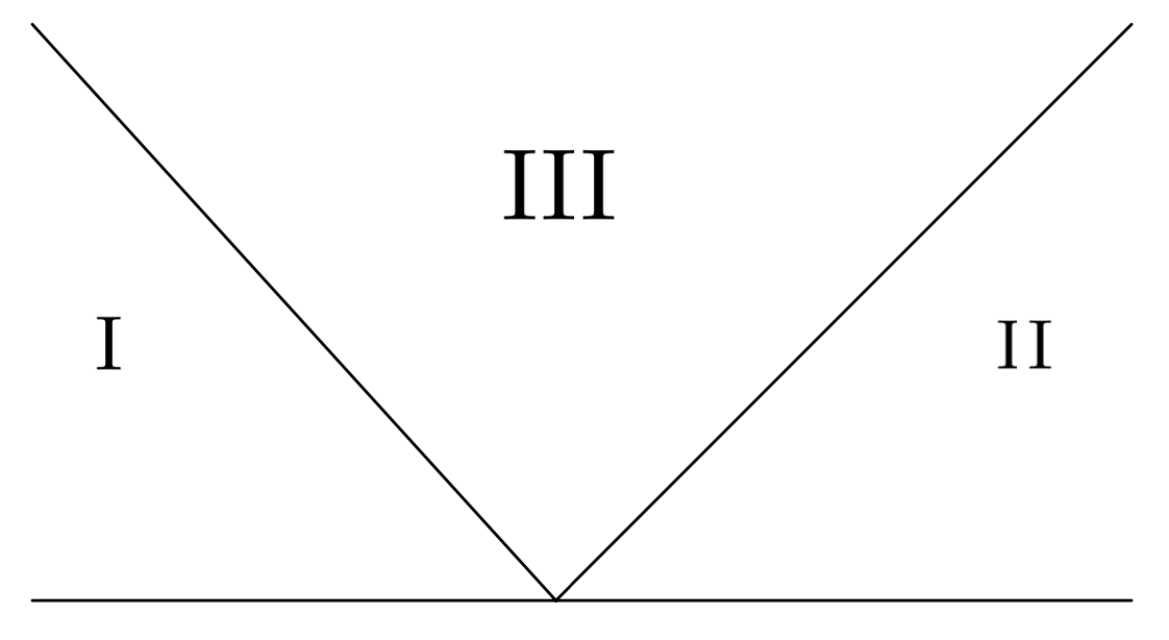}
    \end{minipage}
    \begin{minipage}{.48\textwidth}
        \centering
        \includegraphics[scale=0.3]{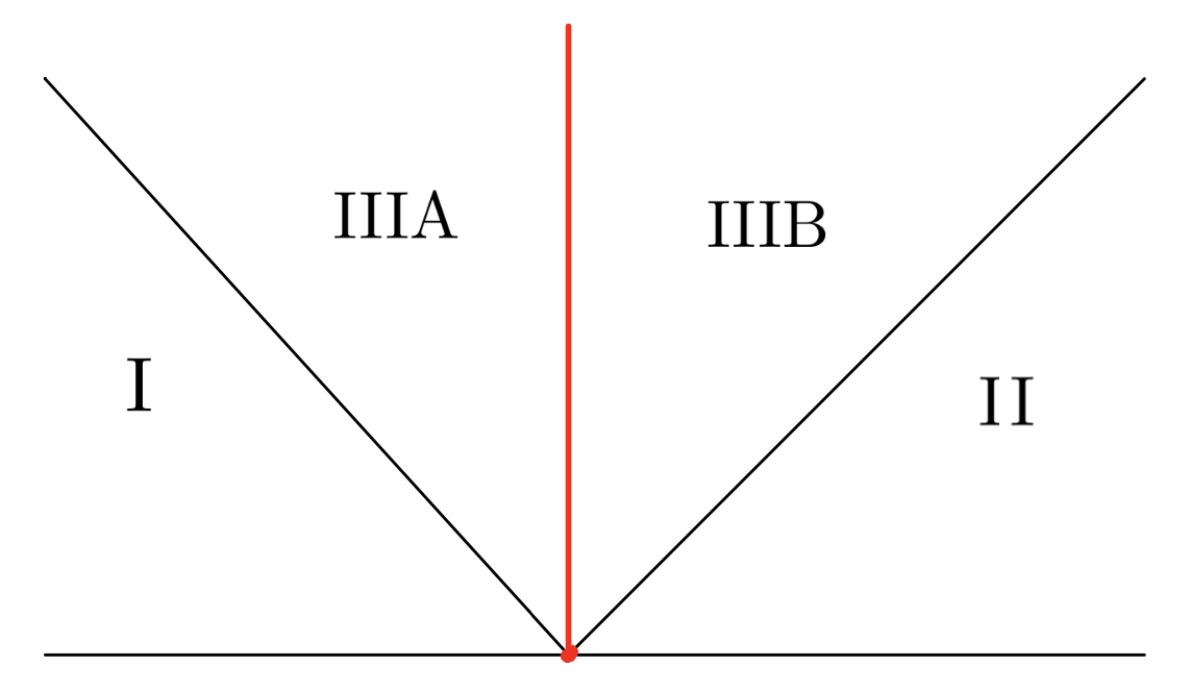}
    \end{minipage}
    \caption{The formation of NESS within the lightcone upon gluing two semi-infinite systems at temperatures $\theta_L$ and $\theta_R$. The NESS forms within the lightcone (III). In the presence of the defect, the NESS region further subdivides into IIIA and IIIB.}
    \label{NESS}
\end{figure}

The NESS is characterized by a constant, time-independent current. In this section, we shall be interested in a NESS with a constant heat current flowing across the interface.   Such a NESS can be prepared using the quench protocol, where two semi-infinite systems at equilibrium temperatures $\theta_L$ and $\theta_R$ are glued at some initial time, say $t=0$. In this case, the NESS forms within the linearly expanding region (Reg III) as shown in the figure \ref{NESS}. Note that in the regions $I$ and $II$ , there is no heat flux, as the systems are at equilibrium, and hence $\left\langle T(u) \right\rangle_{I,II} =\left\langle \bar T(v) \right\rangle_{I,II} = \frac{\pi  c_{L,R} }{12}\theta_{L,R}^2 $. Now as we enter into  Reg III $(z>0,\bar z >0)$ from Reg I $(u>0, v<0)$, $\left\langle T(u) \right\rangle $ remains the same, i.e, $\left\langle T( u) \right\rangle_{NESS}=\left\langle T(u) \right\rangle_{I}= \frac{\pi c_L }{12}\theta_{L}^2 $ whereas $\left\langle \bar T(v) \right\rangle $ has a discrete jump. A similar analysis from the right will simply lead to $\left\langle \bar T(v) \right\rangle_{NESS}= \left\langle \bar T(v) \right\rangle_{II}=  \frac{\pi c_R }{12}\theta_{R}^2  $. Now since  $\left\langle  T( u) \right\rangle_{NESS} \neq \left\langle \bar T(v) \right\rangle_{NESS}$, there in a net heat flux in Reg III,
\begin{equation}
    \nonumber \left\langle  T^{(xt)} \right\rangle_{NESS} = \left(\left\langle T(u) \right\rangle - \left\langle \bar T (v) \right\rangle \right)= \frac{\pi}{12}\left(c_L \theta_L^2 - c_R\theta_R^2 \right),~~~ \text{where}~~ u=x-t, v=x+t,
\end{equation}
which characterizes a NESS.

Let us now introduce the defect at $x=0$. In this case, the NESS region further subdivides into $IIIA$ and $IIIB$ and the chiral energy densities in these regions are given by 
\begin{align}
    \left\langle T(u) \right\rangle_{IIIA} =&\ \frac{\pi }{12}c_L\theta_L^2~,~~~~\left\langle \bar T(v) \right\rangle_{IIIA}= \frac{\pi}{12}\left(c_L(1- \mathcal{T}_L)\theta_L^2+ c_R \mathcal {T}_R\theta_R^2\right)~~ , \label{glue1}\\
     \left\langle \bar T( v) \right\rangle_{IIIB} =&\ \frac{\pi}{12}c_R\theta_R^2~,~~~~\left\langle T(u) \right\rangle_{IIIB}= \frac{\pi}{12}\left(c_L \mathcal {T}_L \theta_L^2+ c_R(1-\mathcal{T}_R)\theta_R^2\right)~~.\label{glue2}
\end{align}
In writing the above, we have assumed the non-absorption of energy at the interface, i.e. $\mathcal R_i=1-\mathcal T_i$, and also the universality of the transmission coefficients irrespective to the nature of the incident excitation as well as the energy carried by it.  Notice that, if we compute the energy flux in any of these NESS regions, we have 
\begin{equation}
     \left\langle  T^{(xt)} \right\rangle_{IIIA} = \left\langle T(u)\right\rangle_{IIIA}- \left\langle \bar T(v)\right\rangle_{IIIA}=\frac{\pi}{12}\left(c_L \mathcal T_L \theta_L^2 - c_R \mathcal T_R \theta_R^2 \right)= \left\langle  T^{(xt)} \right\rangle_{IIIB}. \label{nessflux}
\end{equation}
 The equality must hold by the conformal conditions at the interface, which requires the continuity of $\langle T^{xt} \rangle$. Now when $\theta_L=\theta_R$, the NESS current must vanish. This requires 
\begin{equation}
    c_L\mathcal T_L=c_R\mathcal T_R~~, \label{detailedbalance}
\end{equation}
which is the condition of detailed balance, that must be additionally imposed along with the conditions from conversation conservation. Plugging this condition back in $\eqref{nessflux}$ we have
\begin{equation}
     \left\langle  T^{(xt)} \right\rangle_{IIIA} = \frac{\pi}{12} c_L \mathcal T_L \left(\theta_R^2 - \theta_L^2 \right) .
\end{equation}
This shows that the transmission coefficient can be read off from the Stefan-Boltzmann coefficient. The holographic version of this computation was done in \cite{Bachas:2021tnp}. 

Having discussed the CFT case, let us now turn on the deformation.  Firstly, now the NESS  forms in a region inside (outside) the lightcone of the undeformed theory for $\mu>0 \, (\mu <0)$. Next, the stress tensor of the deformed theory is off-diagonal{\gp, as can be seen from \eqref{defT}}. However, since the off-diagonal piece is completely fixed in terms of the diagonal ones by the trace condition, the gluing does not impose any additional constraint on this term. Also note that, since the detailed balance condition follows from the vanishing of the NESS current and since the current does not depend on the trace, we assume the detailed balance condition \eqref{detailedbalance} to remain true even in the deformed ICFT.  However, a linear matching condition like \eqref{glue1}-\eqref{glue2} across the interface {\gp will} not hold in the deformed theory.  We propose to replace it with a non-linear gluing condition,  that we parametrize in terms of a \textit{transmission function $M(x,y)$} as follows\footnote{\ab{In writing the conditions, we have assumed $c_L=c_R$, which is a simplifying assumption that we shall eventually make in the computations to follow.}}:
\begin{equation} \label{nonlinearmatching}
    \begin{split}
        T_{VV}^A & ={\cal R}^{(0)} \, T_{UU}^A + {\cal T}^{(0)} \, T_{VV}^B- M( T_{UU}^A , T_{VV}^B ) \,, \\
    T_{UU}^B & = {\cal T}^{(0)} \, T_{UU}^A + {\cal R}^{(0)} \, T_{VV}^B+ M( T_{UU}^A , T_{VV}^B ) \,,
    \end{split}
\end{equation}
where $\mathcal T^{(0)}$ is the transmission coefficient in the undeformed ICFT  with $\mathcal R^{(0)}= 1- \mathcal T^{(0)}$ and  $A(B)$ stands for Region $IIIA(B)$. For the folded setup of ICFT, the analogous matching conditions are simply obtained from \eqref{nonlinearmatching} under the exchange  $T_{VV}^B \leftrightarrow T_{UU}^B$.  These relations assure the continuity of $T_{UU}-T_{VV}$ across the interface. So once the continuity of current is imposed, the two matching conditions are equivalent. In that case, one must follow from the other under $A \leftrightarrow B$ exchange, which requires the function $M$ to be completely antisymmetric in terms of the incoming fluxes, i.e. $M(x,y)=-M(y,x)$.

Let us now compute the function $M(x,y)$ for the NESS created in the quench protocol. In this case, the functions $f_i', \bar f'_i, h_i',\bar h'_i, \alpha'_i, \bar \alpha'_i$ appearing in \eqref{pcfloweq}, \eqref{tildemap}, and \eqref{eq:f} are all constants.  We solve the conformal boundary condition in the undeformed theory by taking  $f'_{L} = \frac{M_{L} l+J}{4} \,, f'_{R} = \frac{M_{R} l+J}{4} \,, \bar f'_{L} = \frac{M_{L}l-J}{4}\,, \bar f'_{R} = \frac{M_{R} l -J}{4} $, where the continuity of current fixes $J_L=J_R=J$. The parameters $M_{L,R},J,l$ are the mass, angular momentum, and the AdS radius of the dual solutions in the bulk. For the NESS, we choose $l_L=l_R=l$, that is $c_L=c_R$ as well as $\mu_L=\mu_R=\mu$. The undeformed transmission coefficient is defined by $\frac J l = \frac{{\cal T}^{(0)}}{{\cal R}^{(0)}} \, \frac{M_L-M_R}{2}$.
The equation \eqref{eq:f} yields $h'_{L,R},\bar h'_{L,R}$ and three out of the four parameters of the conformal mapping, $\alpha'_{L,R}, \bar \alpha'_{L,R}$ can be fixed from \eqref{bc1}-\eqref{bc2}. Finally, with all these, we can compute the stress-tensor components in the deformed theory using \eqref{defvev}. One can explicitly check that these vevs satisfy the continuity of $T_{UU}-T_{VV}$ across the interface. 

Let us now solve the matching conditions \eqref{nonlinearmatching}. Since the current is continuous, one of the conditions automatically implies the other. So we look to solve only the first condition
\begin{equation}\label{matchingNESS}
     T_{VV}^A ={\cal R}^{(0)} \, T_{UU}^A + {\cal T}^{(0)} \, T_{UU}^B- M( T_{UU}^A , T_{UU}^B ).
\end{equation}
An exact solution of the above equation is difficult to 
{\gp obtain}, so we solve the equation perturbatively in the vevs. 

 Now, at zeroth order, all the fluxes vanish, and hence $M^{(0,0)}=0$. At linear order, the undeformed matching condition leads to $M^{(1,0)}=0 \, , M^{(0,1)}=0$. The non-trivial corrections to the matching condition appear from second order onward and in the following we report them up to fourth order:
 
 \begin{align}\label{NESSflow}
   M^{(2,0)} &= - M^{(0,2)} = -8\left(1-\mathcal{T}^{(0)}\right)\mathcal{T}^{(0)} \mu \, ,~ M^{(1,1)}= 0\,. \nonumber \\
     M^{(3,0)} &= -M^{(0,3)}= 24 \mathcal{T}^{(0)} \left(2-7\mathcal{T}^{(0)}+5\mathcal{T}^{(0)2}\right) \mu^2 \,,  M^{(2,1)} = -M^{(2,1)}= -8\left(1-\mathcal{T}^{(0)}\right)\mathcal{T}^{(0)2} \mu^2 \, .\nonumber \\ 
      M^{(4,0)} &= -M^{(0,4)}=  384 \mathcal{T}^{(0)2}\left(5-12\mathcal{T}^{(0)} + 7\mathcal{T}^{(0)2}\right) \mu^3 \,,  ~ M^{(2,2)}=0~ \, ,\nonumber \\  M^{(3,1)} &= -M^{(1,3)}= 192\left(1-\mathcal{T}^{(0)2}\right)\mathcal{T}^{(0)2} \mu^3 \, .   \nonumber \\
      \vdots
\end{align}
 We shall reproduce these results from the holographic computation in the following section. {\ab Note that, in case of a NESS with heat current, one can express the incoming fluxes in terms of the temperatures of the two CFTs, namely $f_L'=\frac{\pi c}{12} \theta_L^2,~\bar f_R'=\frac{\pi c}{12} \theta_R^2 $ with $c= 12 \pi l$, thus relating the mass and angular momentum parameters to the temperatures.  Then in equation \eqref{matchingNESS} all the quantities depend upon the two temperatures $\theta_L$ and $\theta_R$ and one can equivalently solve \eqref{matchingNESS} for $M$ perturbatively in the two temperatures. However, in that case, the resulting expression for the expansion coefficients are more complicated {\gp as the $\mu$ dependence appears both from the modified matching conditions, and from the modified relation between the temperature and the vev of the stress-energy tensor.}

\subsection{Scattering state}
We can also implement this procedure in the case of scattering of a monochromatic wave off the defect in the CFT. Let us consider the CFT energy-momentum tensor 
\begin{equation}\label{CFT-excitation}
     f_i(\tilde u) = -i \epsilon \omega {\cal A}_i \, e^{- i \omega \tilde u} \,,  \quad   \bar f_i(\tilde v) = -i \epsilon \omega {\cal B}_i \, e^{i \omega \tilde v} \,, 
\end{equation}
where $i=L,R$. Note that, in this case, we are considering the unfolded setup unlike the folded setup of NESS state. Since the vevs are coordinate-dependent in this case, the pseudoconformal map \eqref{pcfloweq} is difficult to invert. So we shall carry out the analysis perturbatively in the amplitude and the parameter $\epsilon$ is introduced to keep track of the order. To mimic the gravitational scattering setup of \cite{Bachas:2020yxv}, we particularly choose ${\cal A}_L= 1 \, , {\cal A}_R={\cal T}_L^{(0)} {\cal A}_L \, , {\cal B}_L = {\cal R}_L^{(0)} {\cal A}_L \, , {\cal B}_R=0$. This corresponds to a wave of amplitude unity incident on the interface from the left CFT, with the amplitude of the reflected(transmitted) wave being proportional to $\mathcal{R}^{(0)}(\mathcal{T}^{(0)})$. Similarly, for the conformal mappings, we choose the ansatz
\begin{equation}\label{confmapscat}
     \alpha_i( u_i) = \epsilon ~  a_i \, e^{-i \omega  u_i} \,,  \quad   \bar \alpha_i(\tilde v_i) = \epsilon ~  \bar a_i \, e^{i \omega  v_i} \,.
\end{equation}
With \eqref{CFT-excitation} and \eqref{confmapscat}, we first compute the stress-tensor components in $(u,v)$ coordinates from \eqref{eq:f} and integrate them to solve the matching conditions \eqref{bc1}-\eqref{bc2} for $a_L, a_R,$ and $ \bar a_R$, while $\bar a_L$ remains unspecified. Once we have the stress tensor components in $(U,V)$ coordinates using \eqref{defvev}, the continuity of current across the interface at linear order gives
\begin{align}
    J_L-J_R=& \frac{c_L \mu_L \omega^4~ e^{-i \omega \tilde u}\left(6\left(-1 + {\cal R}_L^{(0)} + {\cal T}_L^{(0)}\right) + \left(c_L \mu_L {\cal T}_L^{(0)} +c_R \mu_R\left(-1 + {
    \cal R}_L^{(0)}
    \right)\right)\omega^2\right)}{\left(6+ c_L \mu_L \omega^2\right)\left(6+ c_R \mu_R \omega^2\right)} \epsilon 
\end{align}
So the continuity of current now requires
\begin{align}\label{Jcont}
    \mathcal R_L^{(0)} + \mathcal T _L^{(0)} = 1\,,~~~~  c_L \mu_L=c_R \mu_R .
\end{align}
While the first equation is the energy conservation relation across the conformal interface, the second one, which was trivially satisfied for the NESS, now gives a general constraint upon gluing a pair of deformed CFTs. We shall rederive these constraints in the holographic computations. As of now, akin to the NESS computations, we shall proceed with the simplifying assumption of $c_R=c_L \equiv c \, , \mu_L =\mu_R \equiv \mu$. With the continuity of current established, let us now consider the matching condition
\begin{equation}\label{matchingscat}
      T_{VV}^A ={\cal R}_0 \, T_{UU}^A + {\cal T}_0 \, T_{VV}^B- M( T_{UU}^A , T_{VV}^B ).
\end{equation}
We shall solve the equation perturbatively in $\epsilon$. Since all the vevs are $\mathcal{O}(\epsilon)$, at leading order, we trivially have $M(0,0)=0$. Now as before, one would naively expect that the undeformed matching condition would render the correction at linear order to vanish. However, the key difference with the NESS case is that now the Schwarzian derivative of the conformal mappings leads to non-trivial contribution even at $\mathcal{O}(\epsilon)$, which is also $\bar a_L$-dependent. This dependence, however, disappears once the antisymmetry is imposed, i.e. $M^{(1,0)}=-M^{(0,1)}$. Imposing this, we solve the matching condition \eqref{matchingscat}, which gives
\begin{align}
    M^{(0,1)}= \frac{2 c\left(1- \mathcal{T}_L^{(0)}\right)\mathcal{T}_L^{(0)}\mu \omega^2}{6+c \left(1- 2 \mathcal{T}_L^{(0)} \right)\mu \omega^2}.
\end{align}
The dependence of the correction on the central charge is reminiscent of the fact that this term originates from the Schwarzian derivative, and hence it was absent in the case of NESS. Thus, up to linear order we have 
\begin{align}
    M(T_{UU}^A,T_{VV}^B)=-\left(T_{UU}^A-T_{VV}^B \right) \frac{2 c(1- \mathcal{T}_L^{(0)})\mathcal{T}_L^{(0)}\mu \omega^2}{6+c \left(1- 2 \mathcal{T}_L^{(0)} \right)\mu \omega^2} + \mathcal{O}(\epsilon^2).
\end{align}
 Note that, at linear order, we can absorb this correction to define a $\mu$ dependent transmission coefficient 
\begin{align}
   T_{VV}^A ={\cal R}_L^{(\mu)} T_{UU}^A + {\cal T}_L^{(\mu)} \, T_{UU}^B,
\end{align}
where
\begin{align}\label{Tmu}
    \mathcal{T}^{(\mu)}= \frac{\mathcal T_L^{(0)}(6-c \mu \omega^2)}{6+ c\left(1-2 \mathcal T _L^{(0)}\right)\mu \omega^2} \, ,  ~~~ \mathcal R_L^{(\mu)}=1-\mathcal{T}_L^{(\mu)}.
\end{align}
Note that a perfectly transparent, or a perfectly reflective,  interface remains the same even when it is gluing a pair of deformed CFTs. For semi-transparent interfaces, the transmission coefficient in the deformed theory satisfies the unitarity bound,  i.e., $0 \leq \mathcal T_L^{(\mu)} \leq 1$, for energies $\omega^2 \leq \frac{6}{\mid \mu \mid c}$. 
{\gp For the negative sign of $\mu$ the cutoff on the frequency is reminiscent of the fact that the energy spectrum in the deformed theory on the cylinder becomes complex above a threshold. For the other sign, the reason for the frequency-cutoff is less obvious.}

{\gp We can also consider} 
the more general situation where the excitation of the CFT energy-momentum tensor is a superposition of two modes of the form \eqref{CFT-excitation} with different frequencies. {\gp Remarkably, the result \eqref{Tmu} is still valid if we replace the frequency with the time derivative of the fluxes; more precisely, we write a matching condition that is allowed to depend on the time derivatives of the fluxes; we find that the solution at linear order is given by:
\begin{equation}\label{Mtime}
     T_{VV}^A  ={\cal R}_0 \, T_{UU}^A + {\cal T}_0 \, T_{VV}^B -2 \frac{{\cal R}_0 {\cal T}_0}{{\cal R}_0-{\cal T}_0}\sum_{n=1}^{\infty} \left( \frac{c \mu}{6} ({\cal R}_0- {\cal T}_0)\right)^n \partial_t^{2n} (T_{UU}^A-T_{VV}^B)\,.
\end{equation}


{\gp When the time dependence is monochromatic, \eqref{Mtime} reduces to \eqref{Tmu}.} 
It is also remarkable that in both the single and double-frequency scattering, the coefficients of the reflection function are completely determined by $\mu$ and the parameters of the CFT defect. This can be seen as a partial generalization of the universality property of the ICFT transmission, namely: {\bf the energy transport across the interface depends on the state in a way that is completely specified by the universal function $M$}.  It would be interesting to establish that this holds for a completely generic excitation and to all orders in the perturbation. Of course the universality in the CFT is stronger, in the sense that one can show the independence of the transport on the operators that create the excitation. As we reviewed in section \ref{sec:CFT-universality}, the proof relies on properties of the expansion in conformal blocks of a CFT correlator, so it is not readily  apparent how this could be extended to the $T \bar T$-deformed case. }

\section{Energy transmission in the holographic  NESS}\label{sec:holoNESS}
The universal NESS described in the previous section in the context of two-dimensional CFTs is  characterized by two inequal vevs of the chiral components of the stress-tensor. For holographic CFTs, such a state is described by a black hole given by the metric

\begin{equation}
    ds^2 = \frac{l^2 d\rho^2}{4 \rho^2} + \frac{1}{\rho}\left(du +  \rho \bar{{\mathcal L}} dv\right)\left(dv + \rho {\mathcal L} du\right), \label{rotbtz}
\end{equation}
in the Fefferman-Graham gauge subject to the identification{\footnote{To avoid notational clutter, in this section we shall work in the units $8\pi G =1$.}} $\left\langle T_{uu}\right\rangle =\frac{{\mathcal L}}{l} $, $\left\langle T_{vv}\right\rangle=\frac{\bar {\mathcal L}}{l} $. The metric \eqref{rotbtz} describes a rotating BTZ black hole or a boosted BTZ black hole depending on $u,v$ being compact or non-compact respectively.  Now as discussed in section \ref{sec:floweq}, the deformed geometry can be obtained from \eqref{rotbtz} by inverting the map
\begin{align}
    U= u - \frac{2\mu}{l} \bar {\mathcal L} ~v\, , ~~~V= v- \frac{2 \mu}{l} \mathcal{L}~ u \, .
\end{align}
In BTZ-like coordinates, the deformed geometry takes the form
\begin{align} \label{defrotbtz1}
ds^2_{[\mu]}&\,= \frac{l^2 r^2~ dr^2}{r^4-Ml^2 ~r^2 +\frac{J^2l^2}{4r^2}} - \frac{16(r^2-Ml^2)-8l(J^2 +2M(r^2-Ml^2))\mu-4(r^2-Ml^2)(J^2-M^2l^2)\mu^2}{\left(4+\left(J^2-M^2l^2\right)\mu^2\right)^2}dt^2 \nonumber\\
\nonumber &\ + \frac{4r^2(2-J\mu+Ml\mu)(2+J\mu+Ml\mu)-8J^2l\mu}{\left(4+\left(J^2-M^2l^2\right)\mu^2\right)^2} dx^2+ \frac{4J\left(4l^2 M \mu-8r^2\mu-l^3M^2\mu^2-l(4-J^2\mu^2)\right)}{\left(4+\left(J^2-M^2l^2\right)\mu^2\right)^2}dtdx,\\
\end{align}
which can be obtained from the Fefferman-Graham gauge using the map 
\begin{align}
\nonumber \rho= \frac{r^2 -({\mathcal L}+\bar{\mathcal L})+\sqrt{(r^2 -({\mathcal L}+\bar{\mathcal L}))^2-4{\mathcal L}\bar{\mathcal L}}}{2{\mathcal L}\bar{\mathcal L}}~,~~~~U=x-t~,~~~~V=x+t,~~~~
\end{align}
along with the identification $${\mathcal L}=\frac{l}{4}(Ml+J)~,~~~~\bar{\mathcal L}=\frac{l}{4}(Ml-J) \, ,$$ where  $M,J$ are the mass and angular momentum of the undeformed black hole. Note that, \eqref{defrotbtz1} is also a solution of Einstein's equation similar to its undeformed counterpart. In what follows, we shall derive the energy transmission function for NESS from the gluing of a pair of deformed geometries \eqref{defrotbtz1} with $l_L=l_R=l$, and  $\mu_L=\mu_R=\mu$, across a thin tensile brane with tension $0 \leq \lambda \leq \frac{2}{l}$. This ensures that the worldsheet is locally $AdS_2 $. Now the gluing of the two geometries will be carried out according to the Israel gluing conditions
\begin{align}
    [h_{ab}]=& ~0,\label{israel1}\\
    [K_{ab}]-h_{ab}[K]=& ~\lambda h_{ab}, \label{israel2a}
\end{align}
    where $h_{ab}$ is the induced metric on the hyoersurface, $K_{ab}$ being the extrinsic curvature and $K=h^{ab}K_{ab}$ is its trace. Here $[.]$ implies the jump of the quantity across the membrane.  In what follows, we shall consider the trace-reversed version of \eqref{israel2a} which is given by
    \begin{align}
        K_{ab}^L+K_{ab}^R =- \lambda h_{ab}, \label{israel2b}
    \end{align}
where the superscripts $L,R$ denote the quantities on the left and right of the membrane respectively. In the boundary, quantities with subscript $L(R)$ will correspond to those in the NESS region $IIIA(B)$.

Now consider gluing of two deformed rotating BTZ spacetimes \eqref{defrotbtz1} across a thin brane parametrized by worldsheet coordinates $(\tau,\sigma)$. The embedding of the brane in the two geometries are characterized by six functions $r_{L,R}(\tau,\sigma),~ t_{L,R}(\tau,\sigma)$ and $x_{L,R}(\tau,\sigma)$. The most general stationary embedding that results in a time-independent worldsheet metric is given by 
$$
r_{i}(\tau,\sigma)= R_{i}(\sigma), ~~~ x_{i}(\tau,\sigma)=X_{i}(\sigma)
,~~~t_{i}(\tau,\sigma)= \tau + T_{i}(\sigma),$$
where $i=L,R$. Now the worldsheet theory enjoys reparametrization invariance, using one of which we can set $h_{\tau\tau}=-\sigma$. Then the $\tau\tau$ component of \eqref{israel1}   gives
\begin{equation}
    R_{i}(\sigma)= \sqrt{\frac{\sigma \left(4+ (J_{i}^2 -M_{i}^2 l_{i}^2)\mu^2 \right)^2+4 l_{i} (  2- \mu M_{i}l_{i})\left( M_{i} l_{i}( 2-\mu M_{i}l_{i}) +J_{i}^2  \mu  \right)}{4 \left(2+ (J_{i}-M_{i}l_{i})\mu\right)\left(2- (J_{i}+M_{i}l_{i})\mu \right)}}. \label{rsoln}
\end{equation}
Note that, we are using the same deformation parameters on both sides for simplicity. Now we are left with four functions $T_{L,R}(\sigma)$ and $X_{L,R}(\sigma)$. Out of these four, $T_L+T_R$ is a pure gauge and can be removed using the remaining reparametrization freedom. This leaves us with three physical degrees of freedom to be determined by the remaining of the gluing conditions. Apparently, the system now seems to be overdetermined with five equations left to determine three unknowns. However, thanks to the momentum constraints, $D^{a}K_{ab}-D_{b}K=0$, where $D_{a}$ is the covariant derivative with respect to the induced metric, only one of the three equations in \eqref{israel2b} are independent. These leave us with exactly three equations : i) continuity of $h_{\tau\sigma}$ , ii) continuity of $h_{\sigma\sigma}$, and iii) discontinuity of $K_{\tau\tau}$, for the three unknown which we solve as below. 

Firstly, the continuity of $h_{\tau\sigma}$ gives,
\begin{align}
    \Delta T'(\sigma)= T_L'(\sigma)-T_{R}'(\sigma)=\frac{1}{\sigma}\left[\frac{2 J_{L}(l_L+2\mu\sigma)~X_L'(\sigma)}{\left(2+ (J_L-M_Ll_L)\mu\right)\left(2- (J_L+M_Ll_L)\mu\right)}- L \leftrightarrow R\right].
\end{align}
This leaves us to determine $X_{L,R}(\sigma)$ from the remaining of the two equations, which then completely solves the gluing problem. However, these are ordinary differential equations and an exact analytic solution may be difficult to come up with. So we adapt an alternate strategy following \cite{Bachas:2021tnp}. Firstly, we compute the determinant of the induced metric $h= -\text{det} h_{ab}$ from both sides and invert them to get 
\begin{equation}
    X_{L,R}'(\sigma)= f(h, M_{L,R},J_{L,R},l_{L,R},\mu), \label{xpsoln}
\end{equation}
where we refrain from writing the exact functional dependence for brevity.
Next we consider the extrinsic curvature equations \eqref{israel2a}. Note that, even though there is apparently a single independent equation, which we take to be the $\tau\tau$ equation, mutual consistency between the three equations requires the condition,
\begin{equation}
     \frac{J_{L}}{4+(J_{L}^2 -M_{L}^2l_{L}^2)\mu^2}=- \frac{J_{R}}{4+(J_{R}^2 -M_{R}^2l_{R}^2)\mu^2} . \label{conservation}
\end{equation}
Note that, this equation is simply the continuity of energy flux in the deformed theory $\left\langle T^{xt} \right \rangle_{IIIA}^{[\mu]} = - \left\langle T^{xt} \right \rangle_{IIIB}^{[\mu]} $ across the interface in the folded ICFT setup. Now the $\tau\tau$ component of \eqref{israel2b} gives,
\begin{align}
    &\frac{4\left(2+ (J_{L}-M_{L}l_{L})\mu\right)\left(-2 +(J_{L}+M_{L}l_{L})\mu\right)\left(4R_{L}(\sigma)^4 -4M_{L}l_{L}^2 R_{L}(\sigma)^2+J_{L}^2l_{L}^2\right)X_L'(\sigma)}{l_{L}\left(4+(J_{L}^2 -M_{L}^2l_{L}^2)\mu^2\right)^3 \sqrt{h}} \nonumber \\ &~~~~~~~~~~~~~~~~~~~~~~~~~~~~~~~~~~~~~~~~~~~~~~~~~~~~~~~~~~~~~~~~~~~~~~~~~~~~~~~+ L \leftrightarrow R =- \lambda \sigma, \label{keqn1}
\end{align}
where $R_{L,R}(\sigma)$ are given by \eqref{rsoln}. Now upon substituting \eqref{xpsoln} in \eqref{keqn1}, the latter becomes an algebraic equation for $h$  which after a significant amount of simplification takes the compact form
\begin{equation}
    \sqrt{A_L h -\sigma}+  \sqrt{A_R h -\sigma} = - 2\lambda \sigma \sqrt{h}, 
\end{equation}
where
\begin{align}A_{L,R}= & \frac{4}{l^2}\left[4+ \left(J_{L,R}^2-M^2_{L,R}l^2\right)\mu^2\right)^{-2}\left(4 J_{L,R}^2 l^2 + 4l \left(4 J_{L,R}^2 ~\mu+ M_{L,R}l \left(2 + \left(J_{L,R}- M_{L,R}l\right)\mu\right) \right. \right. \nonumber\\ &
\left. \left. ~~\left(2-\left(J_{L,R}+ M_{L,R}\right)\mu\right)\right)\sigma + \left(4+ \left(J_{L,R}^2- M_{L,R}^2 l^2\right)\mu^2\right)^2\right]. \nonumber
\end{align}
The equation  admits a solution of the form
\begin{equation}
    h= \frac{\lambda^2 \sigma}{A \sigma^2 + 2B\sigma + C}, \label{detinduced}
\end{equation}
where the coefficients $A,B,C$ depend upon $M_{L,R}\, , J_{L,R}\, , l \,, \mu$ as well as the tension $\lambda$.  
\subsection{Extracting the transmission function}
We can recast \eqref{detinduced} as
\begin{equation}
    -\text{det} h_{ab} = \frac{\lambda^2 \sigma}{A (\sigma-\sigma_+)(\sigma-\sigma_-)} , \label{detinduced2}
\end{equation}
with 
\begin{align}
\nonumber \sigma_{\pm}=\frac{-B \pm \sqrt{B^2 -AC}}{A} , 
\end{align}
and
\begin{align}
\nonumber A = ( \lambda_{\text{max}}^2 - \lambda^2)( \lambda^2 - \lambda_{\text{min}}^2), ~~\lambda_{max}=\left(\frac{1}{l_L}+\frac{1}{l_R}\right),~~ \lambda_{min}=\left(\frac{1}{l_R}-\frac{1}{l_L}\right).
\end{align}
Now since $A>0$, it turns out from \eqref{detinduced2} that the worldsheet becomes spacelike as the strings enters into the the ergoregion $\sigma=0$, if $\sigma_+ <0$. So for timelike worldsheet we must have $\sigma_+ \geq 0$. The case with $\sigma_+>0$ corresponds to a turning point of the membrane \cite{Bachas:2021tnp} and we typically do not want that since our discussion concerns about an isolated defect. For the other choice $\sigma_+=0$, the membrane enters into the ergoregion and remains timelike, and one can further show that it never comes out again \cite{Bachas:2021tnp}. This condition amounts to $B>0$ and $C=0$. The expression  for $C$ takes a much simpler form in terms of the variables
\begin{equation}
    e_{L,R}=\frac{2 M_{L,R}l_{L,R}-\mu \left(J_{L,R}^2-M_{L,R}^2l_{L,R}^2\right)}{4+\mu^2\left(J_{L,R}^2-M_{L,R}^2l_{L,R}^2\right)}~, ~~j_{L,R}= \frac{4 J_{L,R}}{4+\mu^2 \left((J_{L,R}^2 -M_{L,R}^2l_{L,R}^2\right)}.
\end{equation} 
Note that $e$ and $j$ are nothing but the energy and current in the deformed theory. In terms of these variables, the condition for energy conservation \eqref{conservation} takes the extremely simple form $j_R=-j_L$.  Taking  this into account, the condition for a timelike worldsheet inside the ergoregion now takes the form
\begin{align}
  & \left(j l \lambda+ 2 e_R (1+e_R \mu)\left(1+j\mu\left(2l\lambda -2j\mu +j^3 \mu^3\right)\right) +   2 e_L  (1+e_L\mu) \left(-1+ j \mu \left(2 j\mu -j^3 \mu^3 + \right.\right.\right. \nonumber \\ & ~~~~~~~~~~~~~~~~~~~~~~~~~~~~~~~~~~~~~~~~~~~~~~~~~~~~~~~~~~~~~~~~~~~~\qquad \left.\left.\left. 2l\lambda(1+ 2 e_R \mu)^2\right)\right)\right)  = 0. \label{ceqn}
\end{align}
Here we have set $l_L=l_R =l$ for simplicity and also used $j_R=-j_L=j$. To gain insight into the equation, we first take the limit $\mu \rightarrow 0$. Then the condition simply boils down to 
\begin{equation}
    M_L-M_R= \lambda J_L \label{undefSB}
\end{equation}
Now to extract Stefan-Boltzmann law from this conditions one simply needs to express the mass parameters in terms of the known vevs of the stress tensor in the boundary, that is 
\begin{equation}
    M_{L,R}= \frac{4}{l}\left\langle T_{--} \right\rangle - \frac{J_{L,R}}{l} = 8\pi^3 \theta_{L,R}^2\mp \frac{J_{L,R}}{l}. \label{undefmap2}
\end{equation}
where we are considering the folded setup of ICFT following \cite{Bachas:2021tnp}, and  we have used the fact that in the region $IIIA(B)$, the right movers are purely thermal at temperature $\theta_{L(R)}$. Using \eqref{undefmap2} back in \eqref{undefSB} along with the relation $J_R=-J_L$, one can extract the transmission coefficient from the resulting Stefan-Boltzmann law $J_L= \frac{\pi c}{6} \mathcal T_L^{(0)} (\theta_L^2 -\theta_R^2)$ with
\begin{align}
    \mathcal T_L^{(0)} =\frac{2}{l}\left(\frac 2 l +  \lambda \right)^{-1}.
\end{align}
However, in the case of deformed ICFT, this is more subtle. To impose the matching conditions, let us first express the energies and the current in terms of the stress tensor components
\begin{align}
    e_{L(R)} &= T_{UU}^{A(B)}+T_{VV}^{A(B)}- 2\mu T_{UV}^{A(B)}, \label{defmass} \\
    j &= T_{UU}^{A}-T_{VV}^{B}.
\end{align}
where $L(R)$ stands for the left(right) half of the NESS region $IIIA(B)$. Firstly, the deformed energies depend on the trace, which, however, can be replaced using the trace relation \eqref{trace-eq}, 
\begin{align}
    T_{UV}^{A(B)}= \frac{1-\sqrt{1+ 16 \mu^2 T_{UU}^{A(B)}T_{VV}^{A(B)}}}{4 \mu},
\end{align}
  with the diagonal components being given by
    \begin{align}
        T_{UU}^{A(B)} &= \frac{\bar {\mathcal L}^{A(B)}}{l^2- 4\mu^2 {\mathcal L}^{A(B)}\bar {\mathcal L}^{A(B)} }\, , \\
        T_{VV}^{A(B)} &= \frac{ {\mathcal L}^{A(B)}}{l^2- 4\mu^2 {\mathcal L}^{A(B)}\bar {\mathcal L}^{A(B)} }\, . \label{TVV}
    \end{align}
Now while $\bar {\mathcal L}^{A(B)}$ correspond to thermal expectation value of the stress tensor in the defromed theory
\begin{align}
    \bar {\mathcal L}^{A(B)}= \frac{1- 4\pi^2 l \mu^2 \theta_{L(R)}^2-\sqrt{1-8\pi^2 l \mu^2 \theta_{L(R)}^2}}{8\pi^2 \mu^2 \theta^2_{L(R)}},
\end{align}
 we substitute ${\mathcal L}^{A(B)}$ for $T_{VV}^{A(B)}$ and $\bar {\mathcal L}^{A(B)}$ by inverting \eqref{TVV}. Finally, we substitute the $T_{VV}^{A,B}$ from the current continuity condition $T_{VV}^B=T_{UU}^B+ T_{UU}^A-T_{VV}^A$ and the matching condition
 \begin{align}
        T_{VV}^A & ={\cal R}^{(0)} \, T_{UU}^A + {\cal T}^{(0)} \, T_{UU}^B- M( T_{UU}^A , T_{UU}^B ) \,.
    \end{align}
  After all these substitutions, we solve \eqref{ceqn}  for the transmission function $M$ perturbatively in the vevs, and we precisely reproduce the flow equation results \eqref{NESSflow}. Note that, this precise matching is highly non-trivial since in section \ref{sec:coefficientsfloweq} we evaluated these coefficients directly from the matching condition, whereas in this section, we evaluated them from the holographic expression \eqref{ceqn}. The matching upto quartic order is strongly indicative of their precise agreement to all orders in vevs.


    \section{Gravitational Scattering in the deformed spacetime}\label{sec:Gravscatt}

    In this section, we set up the gravitational scattering following \cite{Bachas:2020yxv} to extract the reflection and transmission coefficients of a holographic interface gluing a pair of $T\bar T$-deformed CFTs. The scattering state is realized by shooting null waves from the asymptotic null infinity of the left (deformed) CFT and collecting the flux at the future null infinities of the two (deformed) CFTs, as described in section \ref{sec:CFT-universality} {\ab(see Figure \ref{Figure2})}. In the holographic model, this boundary scattering can be realized by gluing two deformed spacetimes with boundary gravitons across a thin tensile string. Now as discussed in \ref{sec:floweq}, these deformed geometries are obtained from the corresponding undeformed geometries with boundary graviton \cite{Bachas:2020yxv}
 by inverting the map \footnote{Note that,  the multiplicative factors in this map is slightly different compared to section \ref{sec:floweq}. This is to relate to the results of \cite{Bachas:2020yxv}.}
 \begin{align} \label{pcfscat}
        U_{L,R}= u_{L,R} -\frac{\mu_{L,R}}{2Gl_{L,R}} \int_0^v \bar {\mathcal L}_{(L,R)}(v_{L,R})~ ~,~~~~V_{L,R}= v_{L,R} -\frac{\mu_{L,R}}{2Gl_{L,R}} \int_0^u \mathcal L_{(L,R)}(u_{L,R}),
\end{align}
where the integrands are related to the vevs of the boundary stress-tensor. However, the vevs in this case being coordinate-dependent, the inversion is non-trivial. So following \cite{Bachas:2020yxv} (and also similar to section \ref{sec:coefficientsfloweq}), we shall treat the vevs as infinitesimal excitation over the ICFT vacuum and invert the map perturbatively, leading to the deformed spacetimes
\begin{align}\label{defgeoscat}
      ds^2_{L,R}= & \frac{l_{L,R}^2}{y_{L,R}^2} dy_{L,R}^2 + \frac{l_{L,R}^2}{y_{L,R}^2} dU_{L,R} dV_{L,R}~ + \nonumber \\ & \epsilon \left(1+ \frac{ \mu_{L,R} l_{L,R}}{2 G y_{L,R}^2 }\right) \left(\mathcal{L}_{L,R}(U_{L,R})dU^2_{L,R} +    \bar{\mathcal L}_{(L,R)}(V_{L,R})dV^2_{L,R} \right)+ \mathcal{O}(\epsilon^2),
    \end{align}
     which satisfies Einstein's equation perturbatively. Note that the perturbative geometries know about the deformation even at $\mathcal{O}(\epsilon)$, although the boundary CFTs acquire a trace at $\mathcal{O}(\epsilon^2)$. At linearized order, even in the deformed theory, the functions  $\mathcal L$ and $\bar{\mathcal L}$ are proportional to the stress-tensor vevs (see Eq. \eqref{defT}). Now following \cite{Bachas:2020yxv}, we choose the following  ansatz for the vevs:
    \begin{align} \label{vevs}
       \mathcal L_L(U_L)= 4 G l_L~  e^{i\omega U_L},~~\bar{\mathcal L}_L(V_L)= 4Gl_L~ \mathcal R_L e^{i\omega V_L}, ~~\mathcal L_R(U_R)= 4G l_R~ \mathcal T_L e^{i\omega U_R},~~\bar{\mathcal L}_R(V_R)= 0.
    \end{align}
   This corresponds to  the scattering of a boundary graviton incident on the interface from the left CFT, with the amplitude of the reflected(transmitted) wave being proportional to the reflection(transmission) coefficients $\mathcal R_L(\mathcal T_L)$. This ansatz is also similar to the one considered in section \ref{sec:coefficientsfloweq}, barring the fact that in this case, the reflection and the transmission coefficients correspond to the deformed ICFT.

    Now, to carry out the bulk gluing, we first relabel $U_{L,R}= \mathcal X_{L,R}-t_{L,R}$, $V_{L,R}= \mathcal X_{L,R}+t_{L,R}$ and follow it up with a rotation in the $(\mathcal X,y)$ plane $$\mathcal X_{L,R}= X_{L,R} \cos{\theta_{L,R}}+ \eta~ Z_{L,R} \sin{\theta_{L,R}}, ~~ y_{L,R}= - \eta~ X_{L,R} \sin{\theta_{L,R}}+ Z_{L,R} \cos{\theta}_{L,R} ,$$
    with $\eta =+1$ for left the CFT and $\eta=-1$ for the right CFT, such that the unperturbed string sits at $X_{L,R}=0$ and its worldsheet can be parametrized by $t_{L,R}=t,~ Z_{L,R}=z$. Next, we look to solve the gluing conditions \eqref{israel1}-\eqref{israel2b}\footnote{For this section, we replace $\lambda $ with $8\pi G \lambda$. } perturbatively in $\epsilon$ to find the string embedding subject to the ansatz
  \begin{align}
     t_{L,R}=t + \epsilon~  \gamma_{L,R}(t,z),~~ Z_{L,R}=z + \epsilon~  \zeta_{L,R}(t,z),~~ 
     X_{L,R}= \epsilon~  \delta_{L,R}(t,z).
  \end{align}
Note that, since the geometries are already perturbative, we will use the zeroth-order embedding in \eqref{vevs}. 

\subsection{Zeroth order}
With a straight string sitting at $X_{L,R}=0$, the worldsheet at leading order is $AdS_2$ and continuity of the induced metric constrains 
\begin{align}\label{gluscat01}
    l_L= l_W \cos{\theta}_L,~~l_R= l_W \cos{\theta}_R,
\end{align}
with $l_W$ being the $AdS_2$ radius. The extrinsic curvature condition further fixes the radius as
\begin{align}\label{gluscat02}
    l_W= \frac{\tan \theta_L + \tan \theta_R}{8 \pi G \lambda}
\end{align}
with $\lambda$ being the brane tension.

\subsection{First order}
Next we solve the gluing conditions at $\mathcal{O}(\epsilon)$  for the six functions $\gamma_{L,R}(t,z)$, $\zeta_{L,R}(t,z)$, and  $\delta_{L,R}(t,z)$. However, we shall exploit the worldsheet reparametrization invariance to set $\zeta_R(t,z)=0, \gamma_R(t,z)=0$, and denote $\zeta_L \equiv \zeta$, and $\gamma_L \equiv \gamma$. This leaves us with four equations to solve for four functions. Also,  the leading order worksheet metric being time-independent, we shall choose the following kind of ansatz for the four perturbations 
\begin{align}
    f(t,z)= f(z) e^{i\omega t}.
\end{align}
These brane fluctuations are induced by the boundary graviton modes. It turns out to be convenient to redefine the functions and solve for 
\begin{align}
    D= \delta_L-\delta_R,~~~ \Delta = \tan \theta_L + \tan\theta_R \delta_R -\xi
\end{align}
along with $\gamma$ and $\xi$. The functions $\Delta,~\xi $, and $\gamma$ get determined from \eqref{israel1} , while \eqref{israel2b} fixes $D$.
The exact solutions are not quite illuminating, so we report them later in appendix \ref{Ap1}. In what follows, we shall analyse the consequences of imposing the Dirichlet boundary condition on these fluctuations. It is worth emphasizing here that since these fluctuations are brane DOFs, we can impose a Dirichlet boundary condition on them compared to a mixed boundary condition for gravitational dofs. Firstly, $D(0)=0$, gives
\begin{align}\label{gluscat11}
    \mathcal R_L+\mathcal T_L=1, ~~~~~ \frac{\mu_L}{\mu_R}= \frac{\sec \theta_L}{\sec \theta_R}.
\end{align}
Replacing the angles with the AdS radii using \eqref{gluscat01} and using the holographic relation  $\frac{3l_{L,R}}{2G}=c_{L,R}$, it is easy to verify that \eqref{gluscat11} reproduce the conditions \eqref{Jcont} obtained from the continuity of current across the interface. We emphasize here that even in the deformed theory, the coefficients obey the usual conservation law.

Next, while $\Delta(0)=0$ is trivially satisfied, $\zeta(0)=0$ and $\gamma(0)=0$ fix the coefficients $a_+(\omega)$ and $a_-(\omega)$ of the homogenous plane-wave solutions. Now, as discussed in \cite{Bachas:2020yxv}, it turns out that the Israel gluing conditions along with the Dirichlet boundary conditions are not sufficient to extract the transmission coefficients and one needs additional conditions. In this context, the authors imposed a no-outgoing-wave condition at the Poincaré horizon, which in turn sets $a_+(\omega)=0$. In the present context, this leads to 
\begin{align}\label{tlmu}
 \nonumber  \mathcal T^{(\mu)}_L=&  2\sec\theta_L \left(  32 \pi G^2 \lambda - \mu_L \omega^2 \sec \theta_R \sin(\theta_L + \theta_R) \right)\left[32 \pi G^2  \lambda \left(\sec \theta_L +\sec \theta_R\right)
-    \right. \\& \left.  \mu_L \omega^2 \sec^2 \theta_R \sin(\theta_L + \theta_R)\left(1- \sec \theta_L \sin(\theta_L + \theta_R)  \right)
+ \left( 32 \pi  G^2  \lambda - \mu_L \omega^2 \right) \left( \tan \theta_L + \tan \theta_R  \right)\right]^{-1},
\end{align}
where we have replaced $\mu_R$ using \eqref{gluscat11} and explicitly introduced the superscript to denote the quantity in the deformed theory. Let us first explore some interesting limits of this expression. Firstly, in the limit of vanishing $\mu_L$, we recover
\begin{equation}\label{T0undefscat}
    \mathcal T^{(0)}_{L}= \frac{2 \sec \theta_L}{\sec \theta_L+\sec \theta_R + \tan \theta_L +\tan \theta_R} =  \frac{2}{l_L} \left(\frac{1}{l_L}+ \frac{1}{l_R}+ 8\pi G \lambda \right)^{-1},
\end{equation}
where in the second step have used \eqref{gluscat01}-\eqref{gluscat02}. This precisely agrees with the undeformed transmission coefficient reported in \cite{Bachas:2020yxv}. Next, we consider the zero-tension limit, which also requires $\theta_L=\theta_R=0$ from \eqref{gluscat01}. In this limit, we have $\mathcal T^{(\mu)}_L=1$, which is expected as the interface becomes transparent when the tension vanishes. Finally, to compare with the flow equation results, we consider the equal central-charge limit, which amounts to setting $ \theta_R = \theta_L$, and hence $\mu_L=\mu_R=\mu$ from \eqref{gluscat11}. This gives
\begin{align}\label{defTmuholo}
    \mathcal T^{(\mu)}_L= \mathcal T_L^{(0)} \frac{4G - \mu \omega^2 l}{4G+l\left(1-2 \mathcal T^{(0)}_L\right)\mu \omega^2},
\end{align}
where we have replaced $\theta_L$ and $\lambda$ with $\mathcal T^{(0)}_L$ using \eqref{T0undefscat}. This  agrees precisely with \eqref{Tmu} with the identification $\frac{3 l}{2 G}=c$. 

As shown in \cite{Bachas:2020yxv}, for an undeformed ICFT, the analysis to linear order in vev is sufficient to unambiguously fix the universal coefficient to all orders in the expansion. However, in the case of an interface in deformed CFT, \eqref{defTmuholo} is expected to receive correction from the non-linear\footnote{Note that, even for the deformed geometries,  the Fefferman-Graham expansion truncates at second order, but the Israel gluing conditions do not.} transmission coefficients resulting from the higher derivatives of the transmission function. To capture this, one needs to carry out the gluing beyond linear order and express the outgoing fluxes in terms of the incoming one using our modified matching condition \eqref{nonlinearmatching}. This would be technically more involved and we leave it for future work.

\section{Conclusions and outlook}\label{sec:conclusion}

Let us summarize our work. In this paper, we have studied energy transport properties of an interface gluing a pair of $T \bar T$-deformed 1+1D holographic CFTs. Gluing of such CFTs must obey the relation $c_L\mu_L = c_R \mu_R$. Our approach has been two-fold. Firstly, we have used the flow equations of $T \bar T$ deformation to compute the vevs of the stress-tensor in the deformed theory. Since the flow equations are state-dependent, for the concreteness of our analysis, we chose two classes of states \textemdash ~ a NESS with a heat current and a scattering state. In this context, we have also proposed a non-linear modification of the usual gluing conditions, captured by a function of the incoming fluxes, which we call the transmission function. The non-linear matching conditions ensure continuity of current across the interface. Furthermore, the self-consistency of the matching conditions under the exchange of the incoming fluxes requires the transmission function to be purely antisymmetric in its arguments. For the scattering state, the non-trivial temporal dependence of the fluxes restricts our analysis to linear order only, wherein the transmission function  can be equivalently recast as a frequency-dependent transmission coefficient. This in turn puts an upper bound on the allowed frequencies following from the unitarity constraints. 

In a parallel approach, we have used the mixed-boundary condition interpretation of the deformation along with the thin-brane model of ICFT,  to reproduce the flow equation results by performing a holographic computation in the bulk dual of these states. Towards that, we have given a rigorous account of gluing a pair of deformed geometries across the brane obeying the standard Israel junction conditions. For the NESS, we have reproduced the transmission function up to quartic order in incoming fluxes, whereas for the scattering state we have precisely reproduced the energy-dependent transmission coefficient from various holographic conditions. In 
{\gp We can see these results as a partial generalization of the universality property of the energy transmission coefficient in the CFT, extended now to the non-linear (for the NESS) or frequency-dependent (for the scattering state) coefficients. As we have discussed, however, we cannot ascertain the universality in the sense of independence from the operator that creates the excitation.}

{\gp Several questions remain to be addressed. 
It would be interesting to have a closed-form expression for the transmission function for the NESS; this can perhaps be done by computing more orders  in the perturbative expansion and attempting a resummation. It would be important to extend our analysis beyond linear order for the scattering state. The computation becomes more involved, and it is likely that the condition of current continuity has to be generalized to account for the non-vanishing trace of the energy-momentum tensor (as was shown in the case of a relevant deformation of a topological defect in \cite{Bajnok:2013waa}). It would also be interesting to see if our results could be reproduced by the methods of generalized hydrodynamics \cite{Castro-Alvaredo:2016cdj,Piroli:2017ltm}, imposing a condition at the interface for the scattering of quasi-particles. The most important question is whether our results can be extended beyond the holographic context to the full quantum theory. Also, it would be worth exploring how does the (modified) energy transmsission properties of the interface affect the density of states of the two deformed CFTs. In particular, it would be interesting to investigate the existence of negative specific  heat in the deformed CFTs  \cite{Barbon:2020amo} in the presence of the interface. This perhaps can most neatly be addressed by putting the theory on a finite volume, which however, would require a pair of interfaces rather than one. Finally, it would be interesting to repeat the holographic computation using conformal boundary condition at the cutoff surface; this prescription has been argued to correspond to coupling the CFT to a timelike Liouville theory and deforming by a Liouville-dressed $T\bar{T}$ operator \cite{Allameh:2025gsa}; the resulting theory is expected to have a better UV behavior, and it would be interesting to see if this would cure the violation of unitarity we found in the high-frequency regime of the transmission coefficient.\footnote{We thank Dionysios Anninos for the suggestion.}}

\begin{acknowledgments}
We thank Takato Yoshimura for collaboration during the initial stages of the work, and Costas Bachas, Chris Herzog, Ayan Mukhopadhyay, and  Gerard Watts  for useful discussions. The research of AB in Greece was partially supported by the European MSCA grant HORIZONMSCA-2022-PF-01-01 and by the H.F.R.I call “Basic research Financing (Horizontal support of all Sciences)” under the National Recovery and Resilience Plan “Greece 2.0” funded by the European Union– NextGenerationEU (H.F.R.I. Project Number: 15384). AB also acknowledges the support from the Ulam Postdoctoral Fellowship from the  Polish National Agency for Academic Exchange (NAWA). AB and GP also acknowledge the support from IFCPAR/CEFIPRA (Project No. 6304-3). GP would like to thank the Isaac Newton Institute for Mathematical Sciences, Cambridge, for support and hospitality during the programme ``Quantum field theory with boundaries, impurities, and defects (BID)", where part of the work on this paper was undertaken. This work was supported by EPSRC grant no EP/R014604/1.
\end{acknowledgments}
\appendix
\section{Exact solutions at linear order gravitational scattering}\label{Ap1}
In this Appendix, we provide the solutions of the Israel matching conditions at linear order of the gravitational scattering discussed in section \ref{sec:Gravscatt}. At zeroth, the continuity of the induced metric gives
\begin{align}
   l_L= l_W \cos{\theta}_L,~~l_R= l_W \cos{\theta}_R,
\end{align}
whereas the discontinuity of the extrinsic curvature fixes the radius of the $AdS_2$ worldsheet in terms of the brane tension and its location
\begin{align}
     l_W= \frac{\tan \theta_L + \tan \theta_R}{8 \pi G \lambda}.
\end{align}
Next, at linear order, we solve the four matching conditions for the brane fluctuations $\gamma(t,z) \, , \zeta(t,z) \, , \Delta(t,z) \, ,$ and $D(t,z)$. We decompose these functions in Fourier modes, and since the worldsheet at leading order is time-independent, we can work at fixed frequencies. The continuity of the induced metric fixes the first three functions, while the last one gets determined from the extrinsic curvature condition. We report the radial solutions below:
\begin{align}
\Delta(z) =& \frac{z \left(\left(e^{-i \omega  z \sin \theta _L} +\mathcal{R}_L ~e^{i \omega  z \sin \theta _L} \right)\sec \theta _L -\mathcal T_L~ e^{i \omega  z \sin \theta _R}~\sec \theta _R\right)}{\omega ^2} +z\left(a_+  e^{i \omega  z} + a_-  e^{-i \omega  z}\right) \, , \nonumber \\
\gamma(z) =& -i~\frac{ \left(e^{-i \omega  z \sin \theta _L} +\mathcal{R}_L ~e^{i \omega  z \sin \theta _L} \right) \sec \theta _L  \left(2  \omega ^2 z^2 \cos^2 \theta_L-4\right) +\mathcal T_L e^{i \omega  z \sin \theta _R}\sec \theta _R  \left(2 \omega ^2 z^2 \cos^2  \theta _R-4\right)}{4 \omega ^3} \nonumber \\
& -i ~\frac{ \left(\mu _L \left(e^{-i \omega  z \sin \theta _L}+\mathcal R_L e^{ i \omega  z \sin \theta _L}\right)-\mu _R~ \mathcal T_L e^{i \omega  z \sin \theta _R}\right)\left(\tan \theta _L+\tan \theta _R\right)}{32 \pi  G^2 \lambda  \omega } +\frac{i \left(a_+ e^{i \omega  z}+a_- e^{-i \omega  z}\right)}{\omega } \, , \nonumber \\
D(z)=& -\frac{i}{4 \omega ^3} \left(\left(\omega ^2 z^2 \cos 2 \theta _L +\omega ^2 z^2+4\right) \left(e^{-i \omega  z \sin \theta _L}-R_L e^{i \omega  z \sin \theta _L}\right)+4 i \omega  z \sin \theta _L \left(e^{-i \omega  z \sin \theta _L} \right.\right. \nonumber \\ & \qquad\left.\left. +R_L e^{i \omega  z \sin \theta _L}-\right)-T_L e^{i \omega  z \sin \theta _R} \left(\omega  z \left(\omega  z \cos 2 \theta _R-4 i \sin \theta _R\right)+\omega ^2 z^2+4\right)\right) -  \nonumber \\ & \frac{i}{32 \pi  G^2 \lambda  \omega }\left(\sin \left(\theta _L+\theta _R\right) \left(\mu _L \sec \theta _R \left(e^{-i \omega  z \sin \theta _L}-R_L e^{i \omega  z \sin \theta _L}\right)-T_L \mu _R \sec \theta _L e^{i \omega  z \sin \theta _R}\right)\right)\, , \nonumber \\
\zeta(z) =& \frac{1}{2 \omega ^3}\left(\mathcal T_L e^{i \omega  z \sin \theta _R} \left(\omega  z \cos \theta _R \left(2-i \omega  z \sin \theta _R\right)-2 i \tan \theta _R\right)-i e^{-i \omega  z \sin \theta _L} \left(2 \tan \theta _L - \right.\right. \nonumber \\ & \qquad\left.\left.    \mathcal R_L e^{2 i \omega  z \sin \theta _L} \left(2 \tan \theta _L+\omega  z \cos \theta _L \left(\omega  z \sin \theta _L+2 i\right)\right)+\omega  z \cos \theta _L\left(\omega  z \sin \theta _L-2 i\right)\right)\right) - \nonumber \\
 & \frac{i \sec \theta _L \sec \theta _R \sin \left(\theta _L+\theta _R\right) \left(\mu _R \sin \theta _R~\mathcal T_L  e^{i \omega  z \sin \theta _R}+\mu _L \sin \theta _L \left(e^{-i \omega  z \sin \theta _L}-\mathcal R_L e^{i \omega  z \sin \theta _L}\right)\right)}{32 \pi  G^2 \lambda \omega } + \nonumber \\
 & \frac{i \left(a_+ e^{i \omega  z} - a_- e^{-i \omega  z}\right)}{\omega } \, .
\end{align}
 Clearly, $\Delta(0)=0$ and $D(0)=0$ gives the constraints \eqref{gluscat11}. Upon solving $\gamma(0)=0$, and $\zeta(0)=0$, we get 
 \begin{align}
     a_{\pm}= & \frac{1}{64 \pi  G^2 \lambda  \omega ^2}\left(  \left(32 \pi  G^2 \lambda  \left(\mathcal T_L-2\right)+\omega ^2 \mu _L \mathcal T_L \sec ^2 \theta _R \sin ^2\left(\theta _L+\theta _R\right)\right)\sec \theta _L+32 \pi  G^2 \lambda  \mathcal T_L \sec \theta _R \right. \nonumber \\& \qquad \left.  -\omega ^2 \mu _L T_L \sec ^2\theta _R \sin \left(\theta _L+\theta _R\right) \pm  \left(32 \pi  G^2 \lambda  \mathcal T_L-\mu_L (\mathcal{T}_L-2) \omega ^2\right)(\tan \theta _L+\tan \theta _R)\right) 
 \end{align}
 Finally, setting $a_+= 0$ gives \eqref{tlmu}.
\bibliographystyle{JHEP}
\bibliography{references}

\end{document}